\documentclass[twocolumn]{aastex63}

\usepackage{graphicx}
\usepackage{courier}
\usepackage{amsmath} % or simply amstext
\usepackage{gensymb}
\usepackage{booktabs}

\received{ }
\revised{ }
\accepted{ }
\submitjournal{ApJS}

\shorttitle{Spatially resolved SED fitting with \texttt{piXedfit}}
\shortauthors{Abdurro'uf et al.}
\graphicspath{{./}{figures/}}
\DeclareRobustCommand{\ion}[2]{%
\relax\ifmmode
\ifx\testbx\f@series
{\mathbf{#1\,\mathsc{#2}}}\else
{\mathrm{#1\,\mathsc{#2}}}\fi
\else\textup{#1\,{\mdseries\textsc{#2}}}%
\fi}

\begin{document}

%\title{Spatially resolving properties of galaxies with \texttt{piXedfit}. I. Description of the code and demonstration of its performance using IllustrisTNG and spatially resolved spectrophotometric data of local galaxies}
\title{Introducing \texttt{piXedfit} - a Spectral Energy Distribution Fitting Code Designed for Resolved Sources}

\correspondingauthor{Abdurro'uf}
\email{abdurrouf@asiaa.sinica.edu.tw}

\author[0000-0002-5258-8761]{Abdurro'uf}
\affiliation{Institute of Astronomy and Astrophysics, Academia Sinica, \\
11F of AS/NTU Astronomy-Mathematics Building, No.1, Sec. 4, Roosevelt Rd, Taipei 10617, Taiwan, R.O.C.}

\author[0000-0001-7146-4687]{Yen-Ting Lin}
\affiliation{Institute of Astronomy and Astrophysics, Academia Sinica, \\
11F of AS/NTU Astronomy-Mathematics Building, No.1, Sec. 4, Roosevelt Rd, Taipei 10617, Taiwan, R.O.C.}

\author[0000-0002-9665-0440]{Po-Feng Wu}
\affiliation{National Astronomical Observatory of Japan, 2-21-1 Osawa, Mitaka, Tokyo 181-8588, Japan} 

\author[0000-0002-2651-1701]{Masayuki Akiyama}
\affiliation{Astronomical Institute, Tohoku University, Aramaki, Aoba, Sendai 980-8578, Japan}

%\collaboration{1}{(AAS Journals Data Scientists collaboration)}

%\author{Butler Burton}
%\affiliation{Leiden University}
%\affiliation{AAS Journals Associate Editor-in-Chief}
%\nocollaboration{1}

%\altaffiliation{AASTeX v6+ programmer}

%\collaboration{1}{(LaTeX collaboration)}

%\affiliation{AAS Director of Publishing}
%\affiliation{American Astronomical Society \\
%1667 K Street NW, Suite 800 \\

%\author{Scott Chernoff}
%\affiliation{IOP Publishing, Washington, DC 20005}



%% provide the collaboration status of a group of authors. These commands RDSPS

\begin{abstract}
We present \verb|piXedfit|, pixelized spectral energy distribution (SED) fitting, a Python package that provides tools for analyzing spatially resolved properties of galaxies using multiband imaging data alone or in combination with integral field spectroscopy (IFS) data. \verb|piXedfit| has six modules that can handle all tasks in the spatially resolved SED fitting. The SED fitting module uses the Bayesian inference technique with two kinds of posteriors sampling methods: Markov Chain Monte Carlo (MCMC) and random densely-sampling of parameter space (RDSPS). We test the performance of the SED fitting module using mock SEDs of simulated galaxies from IllustrisTNG. The SED fitting with both posteriors sampling methods can recover physical properties and star formation histories of the IllustrisTNG galaxies well. We further test the performance of \verb|piXedfit| modules by analyzing 20 galaxies observed by the CALIFA and MaNGA surveys. The data comprises of 12-band imaging data from GALEX, SDSS, 2MASS, and WISE, and the IFS data from CALIFA or MaNGA. \verb|piXedfit| can spatially match (in resolution and sampling) of the imaging and IFS data. By fitting only the photometric SEDs, \verb|piXedfit| can predict the spectral continuum, $\text{D}_{\rm n}4000$, $H_{\alpha}$, and $H_{\beta}$ well. The star formation rate (SFR) derived by \verb|piXedfit| is consistent with that derived from $H_{\alpha}$ emission. The RDSPS method gives equally good fitting results as the MCMC and it is much faster than the MCMC. \verb|piXedfit| is a versatile tool equipped with a parallel computing module for efficient analysis of large datasets, and will be made publicly available\footnote{\url{https://github.com/aabdurrouf/piXedfit}}.                 
\end{abstract}



\section{Introduction} \label{sec:intro}
The accumulated multiwavelength photometric and spectroscopic observations over the past decades have played a crucial role in our current understanding of galaxy formation and evolution. To interpret the multiwavelength data, modeling of the galaxy spectral energy distribution (SED) is required. Motivated by such needs, stellar population synthesis modeling has been systematically developed since the pioneering work by \citet{1972Tinsley} and \citet{1973Searle}. Since then, numerous efforts from various groups have been made to improve the methods \citep{1989Buzzoni, 1993Bruzual, 2003Bruzual, 1998Maraston, 2005Maraston, 2009Conroy, 2009Eldridge}. Recently, extensive developments have been made to include more physical components into the SED modeling, to account for the complexity of the physics underlying the SED of a galaxy. These components include nebular emission \citep[e.g.,][]{1998Ferland, 2013Ferland}, dust emission \citep{2005Burgarella, 2007Draine, 2008daCunha, 2008Groves, 2009Noll, 2017Leja}, dusty torus emission from an active galactic nucleus \citep[AGN; e.g.,][]{2008Nenkova_a, 2012Stalevski}, and synchrotron radio emission \citep[e.g.,][]{2019Boquien}.    

In parallel with the development of the SED modeling, the statistical method for comparison between the observed SED and model SED, the so-called SED fitting, has been extensively developed over the past few decades \citep[see reviews by][]{2011Walcher, 2013Conroy}. Traditionally, SED fitting was considered as an optimization problem, where $\chi^{2}$ minimization technique is used to find a model that best reproduce the observed SED \citep[e.g.,][]{1998Sawicki, 1999Arnouts, 2005CidFernandes, 2009Kriek, 2012Sawicki}. As the number of parameters in the SED modeling becomes higher (due to the incorporation of various physical components, as described above) which introducing more opportunities of degeneracy among the parameters, we see the emergence of the Bayesian inference technique. This technique infers the parameters from posterior probability distributions produced by taking into account the likelihoods of all models. Pioneered by \citet{2003Kauffmann}, the Bayesian framework for SED fitting has been applied widely in the literature \citep[e.g.,][]{2005Burgarella, 2007Salim, 2008daCunha, 2009Noll, 2019Boquien}. Currently, a Bayesian inference with state of the art posteriors sampling technique, such as the Markov Chain Monte Carlo (MCMC) and the nested sampling techniques, has become a standard practice in the SED fitting \citep[e.g.,][]{2011Acquaviva, 2011Serra, 2013Johnson, 2014Han, 2016Chevallard, 2016CalistroRivera, 2017Leja, 2018Carnall,2020Zhou}.     

Despite the fact that galaxies are extended objects, the majority of the studies over the past decades have only utilized their integrated light, particularly for SED fitting; in the case of spectroscopic studies, the integrated spectrum of a galaxy is obtained with the single-fiber spectroscopy over a small diameter of the galaxy's center \citep[e.g., Sloan Digital Sky Survey, SDSS, Galaxy and Mass Assembly survey, GAMA, ][respectively]{2000York, 2009Driver}. These observations have revealed many important evolutionary trends and correlations among physical properties of galaxies that shaped our current understanding of galaxy evolution. 

Despite the huge amount of information obtained from the above surveys, we have not made the full use of the available information, namely the omission of spatially resolved SED with which physical properties of spatial regions in the galaxy can be derived. As spatially extended objects, galaxies have properties that vary across their bodies. The advent of the integral field spectroscopy (IFS) surveys has revolutionized the studies of galaxy formation and evolution: in the local universe, we have SAURON \citep{2002deZeeuw}, $\text{ATLAS}^{\rm 3D}$ \citep{2011Cappellari}, CALIFA \citep{2012Sanchez}, SAMI \citep{2012Croom}, and MaNGA \citep{2015Bundy}; in the high redshifts, KMOS$^{3D}$ \citep{2015Wisnioski} and SINS/zC-SINF \citep{2018ForsterSchreiber}. Thanks to these surveys, spatially resolved properties of galaxies are recently being studied, allowing for a better understanding of the galaxy evolution.   

While the SED fitting technique has been widely applied to the integrated SEDs of galaxies over a wide range of redshifts, its potential for applications to the spatially resolved SEDs has only been explored by a limited number of studies. \citet{1999Abraham} did fitting of spectral synthesis models to spatially resolved multicolor photometry of 32 galaxies at $0.4<z<1$ in the Hubble Deep Field (HDF) to study the ages and evolutionary histories of the stellar populations in the galaxies. \citet{2007Lanyon-Foster,2012Lanyon-Foster} analyzed the pixel-by-pixel multicolor photometry of galaxies at $z<1$ using pixel color-magnitude diagram (pCMDs; the similar method is also implemented by \citealt{1986Bothun}) to study the structural parameters of galaxies across the Hubble sequence. \citet{2009Zibetti} used spatially resolved optical/near infrared (NIR) colors to infer spatially resolved mass-to-light ratios (M/L), which are then multiplied by the surface brightness to obtain the maps of stellar mass surface density ($\Sigma_{*}$) of 9 nearby galaxies. \citet{2012Wuyts, 2013Wuyts} applied the standard SED fitting technique to the spatially resolved broad-band SEDs (from the Hubble Space Telescope, HST) of $0.5<z<2.5$ star-forming galaxies in the GOODS-South field. They used the resulting maps of stellar population properties to analyze the variations in rest-frame color, $\Sigma_{*}$, age, and dust attenuation as a function of galactocentric radius, and measure structural parameters of the galaxies.

Recently, \citet{2015Sorba,2018Sorba} used multiband images covering rest-frame ultraviolet (UV)--optical to conduct pixel-by-pixel SED fitting of 67 nearby galaxies and 1222 galaxies in high redshifts (up to $z\sim 2.5$) to study the systematic effect introduced by the integrated SED fitting on the total stellar mass ($M_{*}$) estimate. By comparing the total $M_{*}$ from summing up the spatially resolved mass estimates with that obtained from the integrated SED fitting (i.e.,~spatially-unresolved $M_{*}$), they found that the $M_{*}$ can be severely underestimated using the integrated SED, especially on star-forming galaxies. They argue that this systematic effect is caused by the outshining effect by young stars, i.e.,~young stars (which have low $\text{M}/\text{L}$) are so bright such that their light dominates the galaxy's SED in the optical wavelengths, thus undermining the contribution from old stars (which have high $\text{M}/\text{L}$)\footnote{However, the discrepancy between the both total $M_{*}$ estimates is not observed by \citet{2012Wuyts} and \citet{2018Smith}. \citet{2018Smith} used synthetic galaxy images covering FUV--FIR that are constructed by performing dust radiative transfer on a 3D hydrodynamical simulation of an isolated disk galaxy.}.

In our previous studies \citep{2017Abdurrouf, 2018Abdurrouf}, we conducted spatially resolved SED fitting of 93 local ($0.01<z<0.02$) and 152 high redshifts ($0.8<z<1.8$) massive disk galaxies to study the evolution of the spatially resolved star formation main sequence (SFMS) and the radial trends of disk growth and quenching. Overall, we found that massive disk galaxies tend to build their stellar masses and quench their star formation activities in the inside-out fashion.      

Until recently, the wide area IFS surveys (mentioned previously) have been mostly targeting local galaxies because such large surveys for high redshifts galaxies are prohibitively expensive. The spatially resolved SED fitting method can serve as a powerful alternative to studying the spatially resolved stellar population properties of galaxies across a wide range of redshifts, as shown by previous studies mentioned above. Some advantages of this method over the IFS surveys are the following: (1) the current and future abundance of high spatial resolution and deep multiband imaging data, particularly those from space missions such as \textit{Euclid}, \textit{JWST}, and \textit{Roman Space telescope}, which allow us to perform this method to a large number of galaxies across wide range of redshifts, (2) the recent developments in SED modeling and fitting methods enable a robust and rapid estimation of galaxy properties, (3) the usage of a single method to study galaxies over a wide range of redshift can reduce systematic biases (which would arise when different methods are used for different redshift) in the study of evolutionary trends of the galaxy properties. Motivated by these, in this study, we develop \verb|piXedfit|, pixelized SED fitting, a Python package that provides a {\it self-contained} set of tools for analyzing spatially resolved properties of galaxies from imaging data as well as the combination of imaging data and IFS data.

The structure of this paper is as follows. We describe the data sets used for the analysis of this paper in Section~\ref{sec:data}. In Section~\ref{sec:piXedfit_architecture}, we explain the \verb|piXedfit| design, including descriptions of 4 out of 6 modules. The description of the SED fitting approach and the 2 modules associated with it is given in Section~\ref{sec:SEDfit_procedure}. In Section~\ref{sec:tetsfit_illustris_TNG}, we test the SED fitting performance of \verb|piXedfit| using mock SEDs of the simulated galaxies from the IllustrisTNG. In Section~\ref{sec:test_with_specphoto_data}, we empirically test \verb|piXedfit| modules using spatially resolved spectrophotometric data of local galaxies. Finally, we summarize the analysis of this paper in Section~\ref{sec:summary}. As sections~\ref{sec:data} to~\ref{sec:SEDfit_procedure} are primarily technical and describing the architecture of \verb|piXedfit|, readers who are more interested in the performance can start from section~\ref{sec:tetsfit_illustris_TNG} while referring to Table~\ref{Tab:parameters}. 

Throughout this paper, the cosmological parameters of $\Omega_{m}=0.3$, $\Omega_{\Lambda}=0.7$, and $H_{0}=70\text{km}\text{s}^{-1}\text{Mpc}^{-1}$, the AB magnitude system, and the \citet{2003Chabrier} initial mass function (IMF) are assumed. 

\section{Data} \label{sec:data}
In the analysis throughout this paper, two kinds of data sets are used: imaging data set ranging from far-ultraviolet (FUV) to near-infrared (NIR) and the IFS data. Each of the data sets is briefly described in the following.
\subsection{Broad-band Imaging Data} \label{subsec:broad_band_images}

\subsubsection{GALEX} \label{sec:GALEX}
The Galaxy Evolution Explorer \citep[GALEX;][]{2005Martin} is a space mission equipped with a $0.5$-m telescope with a field-of-view of $1.13$ $\text{deg}^{2}$, a pixel resolution of $1.5''$, and a point spread function (PSF) full width at half-maximum (FWHM) of $4.2''$ and $5.3''$ in the FUV and near-ultraviolet (NUV) bands (effective wavelengths: $1538.6$ and $2315.7\text{\normalfont\AA}$), respectively. The imaging survey has three modes: all-sky imaging survey (AIS), medium imaging survey (MIS), and deep imaging survey (DIS). The typical integrations per tile of those three survey modes are $200$ s, $1500$ s, and $30000$ s, respectively. The $5\sigma$ limiting magnitudes in FUV (NUV) of those three survey modes are $19.9$ ($20.8$), $22.6$ ($22.7$), and $24.8$ ($24.4$), respectively \citep{2007Morrissey}. In this paper, we use imaging data from the DIS whenever available. Otherwise, imaging data from the MIS is used.        

\subsubsection{SDSS} \label{sec:SDSS}
The SDSS \citep{2000York} and its following surveys are providing the largest dataset combining imaging and spectroscopic data, using a dedicated $2.5$-m telescope at Apache Point Observatory. The imaging survey has five filters ($u$, $g$, $r$, $i$, and $z$) with central wavelengths ranging from $3551$ to $8932 \text{\normalfont\AA}$ and pixel resolution of $0.396''$. The SDSS imaging is 95$\%$ complete to $u=22.0$ mag, $g=22.2$ mag, $r=22.2$ mag, $i=21.3$ mag, and $z=20.5$ mag \citep{2004Abazajian}. The median seeing of all SDSS imaging data is $1.32''$ in the $r$-band \citep[see][]{2011Ross}. 

\subsubsection{2MASS} \label{sec:2MASS}
The Two Micron All Sky Survey \citep[2MASS;][]{2006Skrutskie} is an imaging survey of the whole sky in the NIR. The survey uses two $1.3$-m telescopes, one at Mt. Hopkins, Arizona, United States and the other at Cerro Tololo, Chile. The telescopes observe the sky in $J$ (1.24 $\mu$m), $H$ (1.66 $\mu$m), and $K_{s}$ (2.16 $\mu$m) bands. The image product is resampled to $1.0''\text{ pixel}^{-1}$. The point-source sensitivities at signal-to-noise ratio of S/N=$10$ are: $15.8$, $15.1$, and $14.3$ mag for $J$, $H$, and $K_{s}$, respectively. The seeing is $\sim 2.5-3.5''$ \citep{2006Skrutskie}.         

\subsubsection{WISE} \label{sec:WISE}
The Wide-field Infrared Survey Explorer \citep[WISE;][]{2010Wright} mapped the whole sky in four infrared bands: $3.4$, $4.6$, $12$, and $22 \mu$m ($W1$, $W2$, $W3$, and $W4$, respectively). In this paper, we use the imaging data product from the AllWISE data release. The four wavelength bands ($W1$, $W2$, $W3$, and $W4$) have spatial resolutions PSF FWHM of $6.1''$, $6.4''$, $6.5''$, and $12.0''$, respectively. The spatial sampling of the imaging product in the four wavelength bands is $1.375''\text{ pixel}^{-1}$. WISE achieved $5\sigma$ point source sensitivites better than $0.08$, $0.11$, $1$, and $6.0$ mJy in unconfused regions on the ecliptic in the four bands \citep{2010Wright}. In the analysis of this paper, we use only data in $W1$ and $W2$ bands.    

\subsection{Integral Field Spectroscopy (IFS) Data} \label{sec:IFU_data}

\subsubsection{CALIFA} \label{sec:califa_data}
The Calar Alto Legacy Integral Field Area (CALIFA) survey \citep{2012Sanchez} is an IFS survey designed to obtain spatially resolved spectra of around 600 galaxies in the local universe ($0.005<z<0.03$). The observations were carried out with the Postdam Multi Aperture Spectrograph \citep[PMAS;][]{2005Roth} ---in the PPak configuration--- mounted at the $3.5$-m telescope at the Calar Alto observatory. Each galaxy is observed with two different overlapping setups. The low-resolution setup (V500; $R\sim850$) covers $3745-7500\text{\normalfont\AA}$, while the medium resolution setup (V1200; $R\sim1650$) covers $3400-4840\text{\normalfont\AA}$. The observations with the V500 and V1200 setups reached $3\sigma$ surface brightness limits of $\sim 23.0\text{ mag}\text{ arcsec}^{-2}$ and $\sim 22.7\text{ mag}\text{ arcsec}^{-2}$, respectively \citep{2012Sanchez}. In the analysis of this paper, we use the combined data product so-called COMB data cubes from the DR3 release \citep{2016Sanchez}. The COMB data product is a collection of data cubes that combines the spectra from the two observation setups. The COMB spectra cover $3701-7501\text{\normalfont\AA}$ with the spectral resolution (FWHM) of $6.0\text{\normalfont\AA}$. The mean spatial resolution (PSF FWHM) of the data cube is $\sim2.5''$, with a spatial sampling of $1.0''\text{ spaxel}^{-1}$.        

\subsubsection{MaNGA} \label{sec:manga_data}
Mapping nearby Galaxies at Apache Point Observatory \citep[MaNGA;][]{2015Bundy}, a part of SDSS IV \citep{2017Blanton}, is a wide area IFS survey targeting $\sim 10,000$ local galaxies at $0.01<z<0.15$. The MaNGA hexagonal fiber bundles make use of the BOSS spectrographs \citep{2013Smee}. The observed spectra cover $3600-10,300\text{\normalfont\AA}$ with a spectral resolution of $R\sim 1100-2200$. After dithering, MaNGA data cubes have an effective spatial resolution FWHM of $2.5''$ \citep{2015Law} and spatial sampling of $0.5''\text{ spaxel}^{-1}$. In the analysis of this paper, we use the \verb|LOGCUBE| data cubes from the data reduction pipeline \citep[DRP;][]{2016Law}. The data cubes reach a typical $10\sigma$ limiting continuum surface brightness of $23.5\text{ mag}\text{ arcsec}^{-2}$ in a five-arcsecond-diameter aperture in the $g$ band \citep{2016Law}. Detailed descriptions on the survey design and observing strategy are given in \citet{2015Law}, \citet{2016Yan}, and \citet{2017Wake}.   

%%=============================================%%%
\section{\texttt{piXedfit} design} \label{sec:piXedfit_architecture}
\begin{figure*}[ht]
\centering
\includegraphics[width=0.95\textwidth]{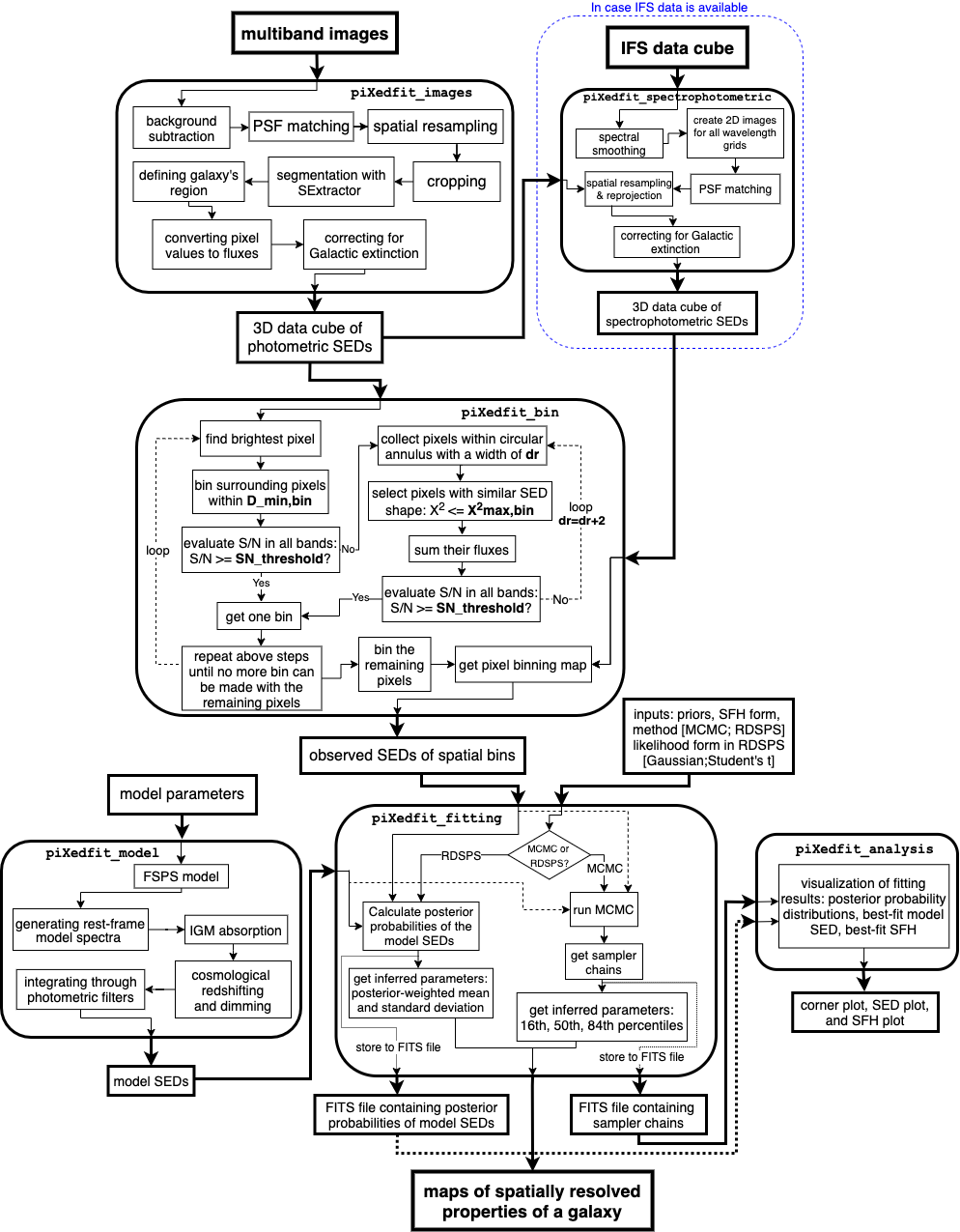}
\caption{The \texttt{piXedfit} design. \texttt{piXedfit} has six modules: (1) \texttt{piXedfit\_images} is for image processing, (2) \texttt{piXedfit\_spectrophotometric} is for spatially matching of multiband imaging data and IFS data to obtain spatially resolved spectrophotometric SEDs of a galaxy, (3) \texttt{piXedfit\_bin} is for pixel binning, (4) \texttt{piXedfit\_model} is for generating model SED, (5) \texttt{piXedfit\_fitting} is for performing SED fitting, and (6) \texttt{piXedfit\_analysis} is for visualizing SED fitting results.}
\label{fig:pixedfit_architecture}
\end{figure*}

\verb|piXedfit| is designed to be modular, and each module can be run independent of each other. Due to its modularity, users can use a particular module in \verb|piXedfit| without the need of using other modules. For instance, it is possible to use the SED fitting module to fit integrated SED of a galaxy (not limited to spatially resolved SED) without the need of using the image processing module. This way \verb|piXedfit| can be beneficial for various applications. Figure~\ref{fig:pixedfit_architecture} shows the design of \verb|piXedfit|. \verb|piXedfit| has six modules: (1) \verb|piXedfit_images| is for image processing, (2) \verb|piXedfit_spectrophotometric| is for spatially matching multiband imaging data with IFS data to obtain spatially resolved spectrophotometric SEDs of a galaxy, (3) \verb|piXedfit_bin| is for pixel binning to maximize $\text{S}/\text{N}$ ratio, (4) \verb|piXedfit_model| is for generating model SEDs, (5) \verb|piXedfit_fitting| is  for performing the SED fitting, and (6) \verb|piXedfit_analysis| is for visualization of fitting results. In this section we describe the first four modules, leaving the last two modules to Section~\ref{sec:SEDfit_procedure}.  

\subsection{\texttt{piXedfit\_images}: Image Processing} \label{sec:piXedfit_images}
In the pixel-by-pixel SED fitting process, it is very important to make sure that the multiband images are all matched to the same spatial resolution and spatial sampling, so that a given pixel represents the same region on the sky in all the images used. Such an image processing task in \verb|piXedfit| is done by the \verb|piXedfit_images| module. The \verb|piXedfit_images| module is a Python scripting module that combines together various image processing functions in \verb|Astropy|\footnote{\url{https://www.astropy.org/}} \citep{2013Astropy}, \verb|Photutils|\footnote{\url{https://photutils.readthedocs.io/en/stable/}} \citep{2019bradley}, and \verb|SExtractor| \citep{1996bertin} such that an image processing task for any combination of imaging data can be done automatically. The user only need to specify a set of photometric bands, the names of input FITS file for the science image associated with each band, the names of input FITS file for the variance image (which is the square root of an uncertainty image) associated with each band, and coordinate (right ascension, RA, and declination, DEC) of the target galaxy. Using a specific function in \verb|piXedfit_images|, the variance image is calculated for each band\footnote{The description on how to estimate the uncertainty of pixel value and derive the variance image are described at \url{https://pixedfit.readthedocs.io/en/latest/list_imaging_data.html}}. The current version of \verb|piXedfit| can perform image processing to the following list of imaging data: GALEX, SDSS, 2MASS, WISE, \textit{Spitzer}, \textit{Herschel}, and Hubble Space Telescope (HST). The workflow of image processing is shown in Figure~\ref{fig:pixedfit_architecture}. In the following, each of the image processing tasks will be described.

\subsubsection{Background Subtraction} \label{sec:background_sub} 
In \verb|piXedfit_images|, the background estimation is done using the \verb|Background2D| function from \verb|Photutils|. The \verb|Background2D| function estimates the background by first dividing an image into certain number of grids and then, for each grid, background level is estimated using the sigma-clipping method. In \verb|piXedfit_images|, grid size is required as an input. 

The background subtraction is only applied to the science images. After the background subtraction process, the background and RMS images are stored into FITS files.     

\subsubsection{PSF Matching} \label{sec:psf_matching}
In order to obtain accurate multiwavelength photometric SED from a set of multiband images, it is important that all the images are brought to the same PSF size. Commonly, PSF matching between two images is done by convolving the higher resolution image (i.e.,~smaller PSF size) with a pre-calculated kernel. The matching kernel between the two PSFs is derived from the ratio of Fourier transforms \citep[see e.g.,][]{2008gordon, 2011aniano}.

Previous studies have constructed convolution kernels for matching the PSFs of imaging data from various telescopes including both space-based and ground-based ones. \citet{2008gordon} constructed convolution kernels for matching the PSFs of the \textit{Spitzer}/IRAC and \textit{Spitzer}/MIPS images\footnote{Convolution kernels are available at \url{https://irsa.ipac.caltech.edu/data/SPITZER/docs/dataanalysistools/tools/contributed/general/convkern/}}. \citet{2011aniano} constructed convolution kernels for matching the PSFs of imaging data from various space-based and ground-based telescopes that includes GALEX, \textit{Spitzer}, WISE, and \textit{Herschel}. Besides that, \citet{2011aniano} also constructed convolution kernels for some analytical PSFs that includes Gaussian, sum of Gaussians, and Moffat\footnote{PSFs and convolution kernels are available at \url{https://www.astro.princeton.edu/~ganiano/Kernels.html}}. The analytical PSF forms are expected to represent the net (i.e.,~effective) PSFs of ground-based telescopes.

We use convolution kernels from \citet{2011aniano} for the PSF matching process in the \verb|piXedfit_images| module. Since the PSFs of SDSS and 2MASS are not explicitly covered in the list of PSFs analyzed by \citet{2011aniano}, to find the analytical PSFs representative of those imaging data, we construct empirical PSFs of the 5 SDSS bands and 3 2MASS bands, then compare them with the analytical PSFs of \citet{2011aniano}. We present this analysis in Appendix~\ref{apdx:comp_empPSFs_aniano2011}. In short, we find that the empirical PSFs of SDSS $u$, $g$, and $r$ bands are best represented by double Gaussian with FWHM of $1.5''$, while the other bands (i.e.,~$i$ and $z$) are best represented by double Gaussian with FWHM of $1.0''$. The two Gaussian components have a fix center, the relative weights of $0.9$ and $0.1$, and the FWHM of the second component is twice that of the first \citep{2011aniano}. For 2MASS, all the three bands ($J$, $H$, and $K_{s}$) are best represented by Gaussian with FWHM of $3.5''$. For consistency, we use those analytical PSFs to represent the PSFs of SDSS and 2MASS and use the convolution kernels associated with them whenever needed\footnote{More information on the kernels and demonstration of their performaces can be seen at \url{https://pixedfit.readthedocs.io/en/latest/list_kernels_psf.html}}. 

In \verb|piXedfit_images|, the convolution of an image with a kernel is done using the \verb|convolve_fft| function in \verb|Astropy|. Before convolving an image with a kernel, the kernel should be spatially resampled to match the spatial sampling of the image, which is done using the \verb|resize_psf| function in \verb|Photutils|. Originally, the kernels provided by \citet{2011aniano} are all resampled to $0.25''\text{ pixel}^{-1}$. The PSF matching process is done to both science images and variance images.  

\subsubsection{Spatial Resampling and Reprojection} \label{sec:spatial_reg}
After the PSF matching, all images are brought to a uniform spatial sampling and reprojection. The final spatial sampling is chosen to be the lowest spatial sampling (i.e.,~largest pixel size) among the imaging data being analyzed. The spatial resampling and reprojection task in \verb|piXedfit_images| is done using the \verb|reproject_exact| function from the \verb|reproject| package \citep{2018robitaille}. \verb|reproject_exact| reprojects an image to a new projection using the flux-conserving spherical polygon intersection method. Because the reprojection basically includes regridding and interpolation, the pixel value of the image should be in a surface brightness unit, not in a flux unit. Therefore before reprojection and resampling, the images are converted into surface brightness whenever needed. If the original unit of an image is in flux, it will be reconverted to flux unit after the resampling process.             

The next step is cropping around the target galaxy. This is done using the \verb|wcs_world2pix| and \verb|Cutout2D| functions available in \verb|Astropy|. The size of the final cropped images, which retain correct WCS information, can be defined by the user. The spatial resampling, reprojection, and cropping are done to the science images and the variance images.

\subsubsection{Image Segmentation and Defining Galaxy's Region of Interest} \label{sec:define_gal_region}
In \verb|piXedfit_images|, image segmentation using \texttt{SExtractor} is done to obtain an initial estimate for the region\footnote{As it is often times difficult to define the boundary of a galaxy, here we refer to the region of the target galaxy to be fit simply as the ``region'' of the galaxy.} of the target galaxy. The segmentation is done in all imaging bands (only the science images), then segmentation maps from all bands are merged (i.e.,~combined) to get a single segmentation map from which the galaxy's region will be determined. Due to the emergence of the background noise, the segmentation map of a galaxy can have an irregular (i.e.,~filamentary) structure at the outskirt. To remove such feature, an elliptical aperture cropping is applied to the galaxy's segmentation region. Ellipticity, position angle, and maximum radius (along the semi-major axis) for the elliptical aperture cropping can be specified when providing input to the \verb|piXedfit_images| module. If those parameters are not provided by the user, elliptical isophote fitting will be done to the final stamp image of a band around the middle of the rest-frame optical (such as $r$ band) using the \verb|Ellipse| class in \verb|Photutils|. In \verb|Ellipse|, the isophotes in the galaxy's image are measured using an iterative method described in \citet{1987jedrzejewski}. From the set of isophotes (as a function of radius) produced by \verb|Ellipse|, the ellipse closest to the desired maximum radius is chosen.

\subsubsection{Extracting SEDs of Pixels} \label{sec:extract_pix_SED}
The tasks described above give the final stamps of reduced science and variance images, and the pixel coordinates associated with the galaxy's region of interest. The next step is calculating fluxes and flux uncertainties of pixels within the galaxy's region in the multiband images. The end product of this process is the photometric SED of every pixel of interest. The conversion of pixel value into flux density unit of $\text{erg }\text{s}^{-1}\text{cm}^{-2}\text{\normalfont\AA}^{-1}$ (which is the default flux unit of data product produced by \verb|piXedfit_images|) depends on the unit of the pixel value in the original image. The flux uncertainty of a pixel is obtained by first taking square root of the pixel value in the variance image then convert it into the flux density unit. 

The pixel values of the imaging data used in our analysis have a variety of units. To convert the pixel value of an image to flux density in $\text{erg }\text{s}^{-1}\text{cm}^{-2}\text{\normalfont\AA}^{-1}$ and estimate the uncertainty of the pixel value, we follow the relevant information from the literature and documentation files from the survey's website from which the imaging data were obtained. The variance images associated with the science images that are input to \verb|piXedfit_images| (see Section~\ref{sec:piXedfit_images}) are constructed following that information\footnote{The unit of pixel value in imaging data that can be analyzed with the current version of \texttt{piXedfit}, and how to convert the pixel value into flux and estimate the flux uncertainty are described at \url{https://pixedfit.readthedocs.io/en/latest/list_imaging_data.html}}.

The next step is to correct the pixel-wise SEDs for the foreground Galactic dust extinction. For this, we estimate $E(B-V)$ from the reddening ($A_{\lambda}$) in the SDSS bands, obtained from the NASA/IPAC Extragalactic Database (NED)\footnote{\url{https://ned.ipac.caltech.edu/}} which is based on the map by \citet{2011Schlafly}, recalibration from \citet{1998Schlegel}. Then we use the \citet{1999Fitzpatrick} with $R_{V}=3.1$ dust reddening law to correct for the foreground Galactic extinction. The final step in the image processing is to crop regions associated with foreground stars. This step is only done if bright stars are found within the galaxy's region of interest. In the current version of the \verb|piXedfit_images| module, this step is done manually using a specific function. The user only need to input central coordinate and an estimate of the radius (in pixels) of each star.       

\begin{figure*}[ht]
\centering
\includegraphics[width=0.7\textwidth]{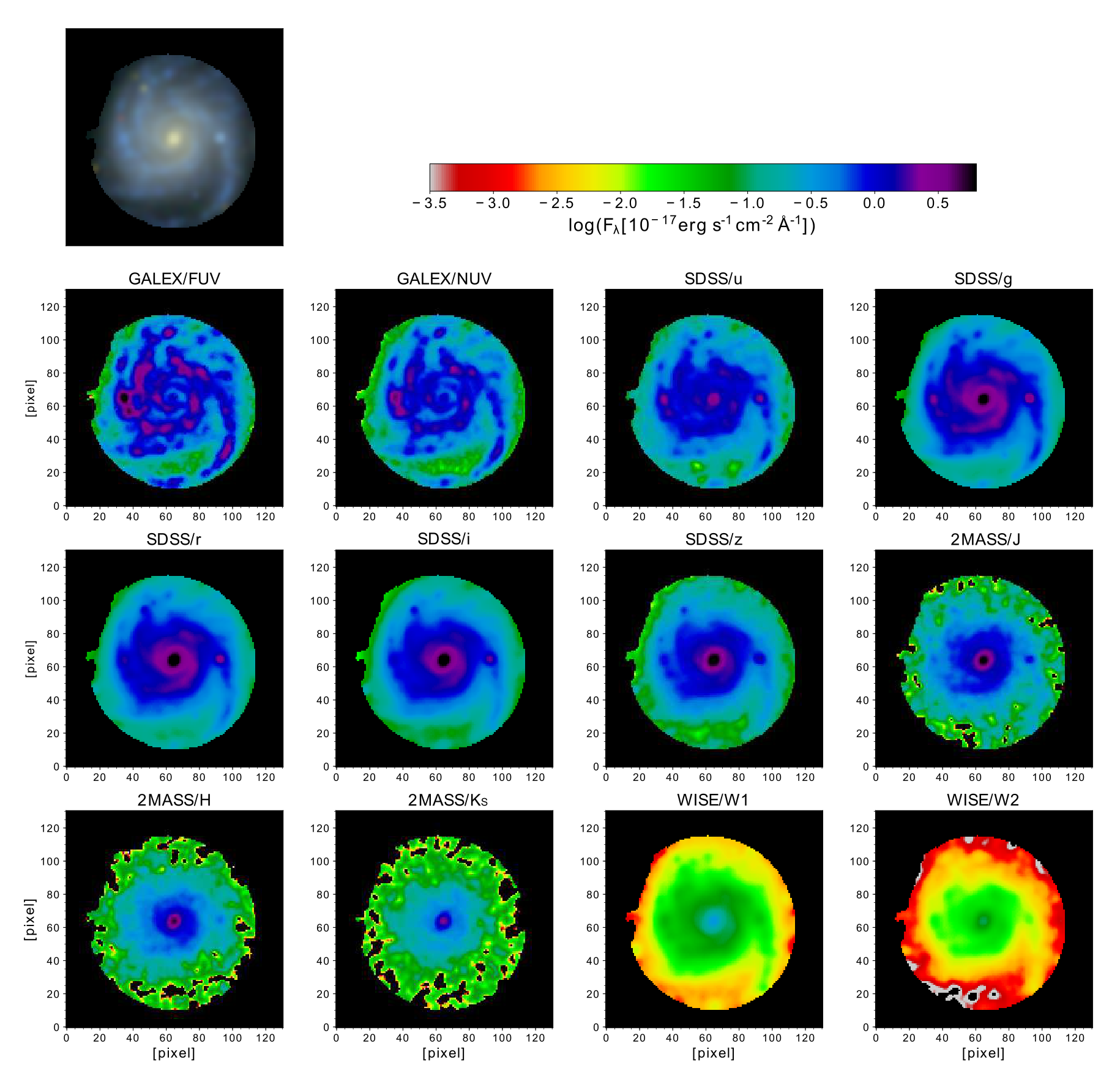}
\caption{Example of the maps of multiband fluxes produced by the \texttt{piXedfit\_images} module. The target galaxy in this example is the NGC 309. The left most panel in the first row shows $gri$ composite image. The 12 panels in the second to fourth row show maps of flux at 12 wavelength bands from the FUV to $W2$. The 12-band images are brought to the spatial resolution of the $W2$ band and the spatial sampling of the FUV/NUV band. The missing pixels in the outskirt of 2MASS bands and $W2$ are caused by the negative fluxes, which do not appear in the logarithmic plot.}
\label{fig:maps_multiband_fluxes}
\end{figure*}

The derived maps of fluxes and flux uncertainties (in multiple photometric bands) of the target galaxy are then saved into one multi-extension FITS file. Figure~\ref{fig:maps_multiband_fluxes} shows an example of the maps of multiband fluxes produced by the \verb|piXedfit_images| module. The target galaxy for this example is NGC 309. Imaging data over 12 bands ranging from the FUV to $W2$ are used to obtain the spatially resolved SED data cube. As can be seen from the fluxes maps, the 2MASS bands are the shallowest among the photometric bands used in the analysis. This is the reason we add the WISE bands ($W1$ and $W2$, which are deeper than the 2MASS bands) although their spatial resolution is lower than UV and optical bands. The inclusion of the WISE bands can provide stronger constraint in the NIR regime.  
 
\subsection{\texttt{piXedfit\_spectrophotometric}: Extracting Spatially Resolved Spectrophotometric SEDs of a Galaxy} \label{sec:piXedfit_spectrophotometric}
In the analyses of the integrated SED of a galaxy (i.e.,~treating the galaxy as one object), there have been several attemps in combining rest-frame optical spectra (particularly covering $4000\text{\normalfont\AA}$ break) and broad-band photometry covering wider wavelength range into a so-called spectrophotometric SED fitting \citep[see e.g.,][]{2014Newman, 2018Dressler, 2019Morishita, 2020Abramson, 2020Chen}. By combining the rest-frame optical spectrum and the broad-band photometry, it is expected that the constraining power in the SED fitting can be enhanced and potentially break the existing degeneracies among the parameters in the fitting process.             

The availability of the FUV--NIR broad-band imaging and the IFS datasets for local galaxies (thanks to CALIFA, MaNGA, and SAMI surveys) give us opportunities to conduct the spatially resolved spectrophotometric SED analyses. However, for a self-consistent analysis we need to spatially match (in spatial resolution and sampling) the broad-band imaging and IFS datasets. \verb|piXedfit| provides a new capability of combining the broad-band imaging data with the IFS data to obtain spatially resolved spectrophotometric SEDs of a galaxy. The tasks featuring this process is in the module \verb|piXedfit_spectrophotometric|. As for the current version, the \verb|piXedfit_spectrophotometric| module can only analyze the combination of broad-band imaging data from the GALEX, SDSS, 2MASS, and WISE, and the IFS data from the CALIFA/COMB and MaNGA/DRP. The final product of this module is a data cube that contains spatially-matched pixel-wise spectrophotometric SEDs of a galaxy. \textit{Our analysis presented here is the first attempt of this kind}. 

To spatially match the three dimensional IFS data with the broad-band imaging data, first, a two dimensional image (i.e.,~the imaging layer) of every wavelength grid in the IFS data is made. Before creating images out of the IFS data, the spectra are smoothed by convolving them with a Gaussian kernel with a sigma value following that of the spectral resolution of the IFS data ($\sim 2.6\text{\normalfont\AA}$ for CALIFA and $\sim 3.5\text{\normalfont\AA}$ for MaNGA, adopted the median value of the spectral resolution across the whole wavelength range). After two dimensional images are created out of the IFS data, the PSF matching and spatial resampling are done to each image, in the same way as processing a broad-band image. In case of matching IFS data from MaNGA or CALIFA with the 12-band imaging data from the FUV to the $W2$, the final product has the spatial resolution of the $W2$ ($6.37''$ FWHM) and the spatial sampling of the FUV/NUV ($1.5''\text{ pixel}^{-1}$).  

The PSF matching for an imaging layer is done by convolving the image with a pre-calculated kernel. Since the effective PSFs of MaNGA and CALIFA have FWHM of $2.5''$, we use corresponding convolution kernel from \citet{2011aniano}. The convolution kernel was created for matching a Gaussian PSF with FWHM of $2.5''$ to the PSF size of $W2$. We have compared the reconstructed PSFs of MaNGA DRP data cube in the $g$, $r$, $i$, and $z$ bands (provided in the FITS file containing the data cube of one galaxy) with the Gaussian PSF with FWHM of $2.5''$ from \citet{2011aniano}. The MaNGA empirical PSFs match well with the Gaussian PSF in all these bands. 

After PSF matching, all the imaging layers are spatially resampled and reprojected to match the spatial sampling and projection of the broad-band imaging data cube produced by \verb|piXedfit_images|. This task is done in the same way as that for the images processing, described in Section~\ref{sec:spatial_reg}. The next step is correcting the spatially resolved spectra for the foreground Galactic dust extinction. This step is only done for the MaNGA data cubes \citep{2016Law}, as such a correction has been applied to the CALIFA cubes \citep{2016Sanchez}. For this task, we use the $E(B-V)$ value obtained from the header (keyword:\verb|EBVGAL|) of the MaNGA DRP FITS file and then apply the dust extinction correction adopting the \citet{1999Fitzpatrick} reddening law with $R_{V}=3.1$.     

We have found that, the normalization of the IFS spectra and the photometric SEDs are often slightly offset from each other. There appears to be no general patterns for the flux offsets. In addition to flux offsets that vary across wavelength in an SED of a pixel, there are also variations of the flux offset spatially. To get a simplified pattern of the variation of the flux offsets, first, we reconstruct $g$, $r$, and $i$ ($g$ and $r$) images from the post-processed IFS data from MaNGA (CALIFA) by convolving them with broadband filters. We then compare the reconstructed images with the real images. For MaNGA, the mean $\log(f_{\rm obs}/f_{\rm recons})$ in $g$, $r$, and $i$ are $-0.014\pm 0.085$, $-0.044\pm 0.092$, and $-0.032\pm 0.090$, respectively. For CALIFA, the $\log(f_{\rm obs}/f_{\rm recons})$ in $g$ and $r$ are $-0.065\pm 0.141$ and $-0.043\pm 0.133$, respectively, where $f_{\rm obs}$ and $f_{\rm recons}$ are flux from real image and the reconstructed image, respectively. These values are derived using a sample of 20 galaxies that will be used in the analysis of Section~\ref{sec:test_with_specphoto_data}.

The mismatch between spectrum and photometric SED can be caused by at least two factors: systematics in the data processing (PSF matching, spatial resampling, reprojection, etc.) of the broad-band imaging data and the IFS data, and the uncertainty in the flux calibration of the photometric and the IFS data. For detailed descriptions on the flux calibration in the MaNGA and CALIFA surveys, please refer to \citet{2016Yan} and \citet{2015Garcia-Benito}, respectively. 

\begin{figure}[ht]
\centering
\includegraphics[width=0.5\textwidth]{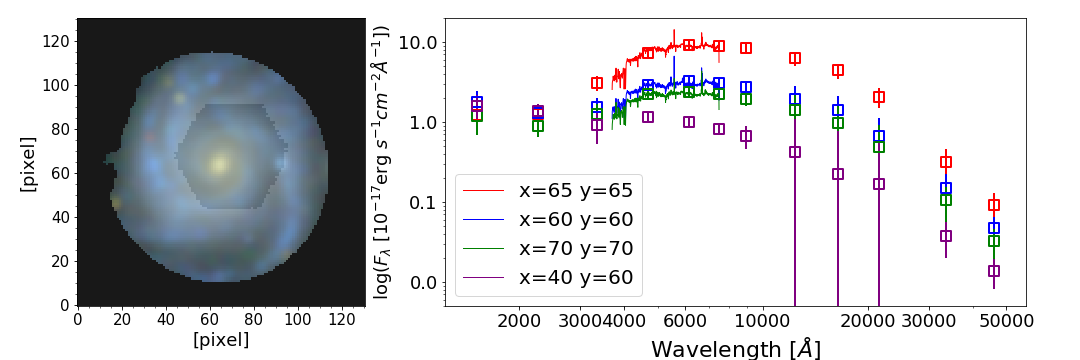}
\includegraphics[width=0.5\textwidth]{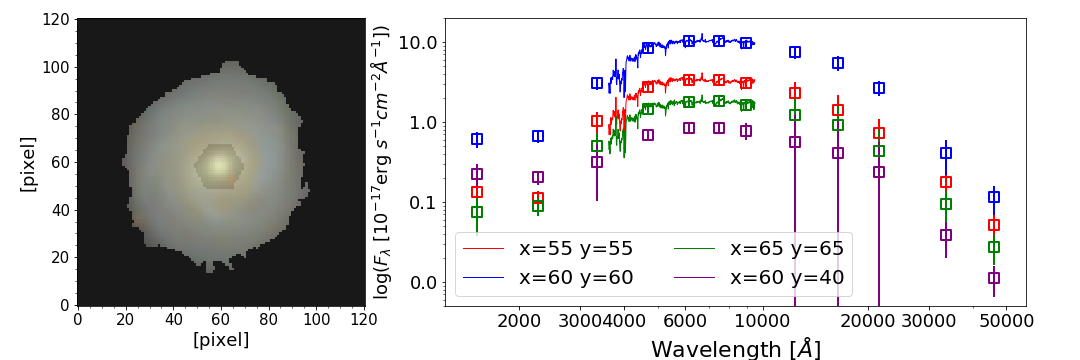}
\caption{Examples of the spectrophotometric SED data cubes obtained with the \texttt{piXedfit\_spectrophotometric} module. The two galaxies are: NGC 309 (first row) which is observed by the CALIFA survey, and a galaxy with PLATE-IFU:8934-12702 (bottom row) observed by the MaNGA survey. The region covered by the IFU fiber bundle is plotted transparently on top of the $gri$ composite image (left panel in each row). In each row, the right panel shows SEDs of 4 randomly chosen pixels --- three spectrophotometric SEDs and one photometric SED (shown by the purple colored points).}
\label{fig:specphotoSED_califa_manga1}
\end{figure}

In order to overcome the photometry--spectroscopy offset, we multiply the spectrum with a wavelength-dependent smooth factor obtained from a third-order Legendre polynomial function fit such that the spectrum normalization become consistent with the normalization of the photometric SED. The polynomial order of $3$ is low enough to prevent the introducing of spectral breaks or artificial features to the spectrum. To find the smooth multiplicative factor, we first obtain a model spectrum that best describes the photometric SED using a $\chi^{2}$ minimization technique applied to a set of pre-calculated model SEDs (to be described in Section~\ref{sec:random_uniform}), then fit a third-order Legendre polynomial to the ratio between the model spectrum and the observed (IFS) spectrum.   
This method adopts the typical technique used in the spectrum fitting that uses multiplicative polynomial function of a certain order ($\sim 2-8$) to make a model spectrum template fit the overall spectral shape of the observed spectrum \citep[see e.g.,][]{2000Kelson, 2009Koleva, 2004Emsellem, 2014Newman, 2017Cappellari, 2019Westfall, 2019Belfiore}. Figure~\ref{fig:specphotoSED_califa_manga1} shows examples of spectrophotometric SED data cubes of the galaxy NGC 309, which is observed by the CALIFA survey (first row), and another galaxy, PLATE-IFU:8934-12702, observed by the MaNGA survey (second row). Regions in the galaxies that are covered by the IFU fiber bundle are shown by the transparent hexagonal regions overlaid on top of the $gri$ composite images (left panel in each row). Outside of these regions, we still have spatially resolved broad-band photometry data. In each row, the right panel shows SEDs of 4 randomly chosen pixels --- three spectrophotometric SEDs and one photometric SED. The $gri$ composite images are made using the \verb|make_lupton_rgb| function in \texttt{Astropy} \citep{2004Lupton}.   

\subsection{\texttt{piXedfit\_bin}: Pixel Binning} \label{sec:piXedfit_bin}
In most cases, fluxes measured in individual pixels have a low $\text{S}/\text{N}$ ratio. It is also common to find pixels with missing or negative fluxes. In order to get an accurate inference of the parameters in the SED fitting, typically one needs an observed SED with sufficient $\text{S}/\text{N}$ ratio. For this reason, we do not apply the SED fitting analysis to pixel-wise SED. Instead, we bin the data locally before conducting further analysis to the data. 

Previous studies have applied pixel binning in spatially resolved SED fitting analysis \citep[e.g.,][]{2013Wuyts, 2019Belfiore, 2018Sanchez}. A popular pixel binning scheme is the Voronoi binning by \citet{2003Cappellari}, who showed that, with the Voronoi tessellation technique, the bins can be made as `compact' as possible, no overlapping with each other, and having similar $\text{S}/\text{N}$ ratio (in a particular band).

In \citet{2017Abdurrouf}, we developed a new pixel binning scheme that takes into account of the similarity in the SED shape among pixels. This new criteria is important especially for the spatially resolved SED fitting analyses, because it is expected to preserve any important information from the SED at the pixel scale. While pixel binning is done to achieve a certain minimum S/N, at the cost of degrading the spatial resolution, we can still preserve important information in the SED at the pixel scale with this binning scheme. In the conventional pixel binning schemes that do not consider the similarity of the SED shape, it is possible that neighboring pixels which have different SED shapes (likely having different properties) are binned together. This could smooth out the spatial variation of the stellar population properties.

\verb|piXedfit_bin| is a module designed for performing such a binning scheme, and is built upon what was developed in \citet{2017Abdurrouf}. There are four requirements in the pixel binning scheme: (1) proximity, such that only neighboring connected pixels are binned together, (2) similarity of SED shape, (3) S/N threshold in each band, and (4) smallest diameter of a bin ($D_{\rm min,bin}$ in pixel). The last requirement is a new parameter introduced with the current version of \verb|piXedfit_bin|. This parameter prevents the binning process from picking a single bright pixel as a bin. In some cases, a single bright pixel (typically around the central region) can exceed the S/N threshold such that further binning with other pixels is not needed. The smallest diameter of the bin can be thought of as the FWHM of the PSF although the user is free to define the diameter.

The pixel binning scheme adopted in \verb|piXedfit_bin| is a simple empirical one. Briefly speaking, a spatial bin is obtained by first selecting a brightest pixel in a reference band which is defined by the user (a band around the middle of the rest-frame optical regime is recommended, e.g.,~the $r$ band). Then pixels enclosed within a diameter of $D_{\rm min,bin}$ from the brightest pixel are joined together and the total S/N of the bin (in each band) is checked. If the total S/N in each band is higher than the S/N threshold, the bin size is not expanded and the first bin is established. Otherwise, the bin's radius is increased by $dr=2$ pixels and pixels within the new annulus are examined to see if they have a similar SED shape as the brightest pixel. Pixels that have similar SED shape are added into the bin and the total S/N at each band is checked. If the total S/N in each band is above the S/N threshold, the expansion of the bin is terminated. Otherwise, the expansion is continued until the S/N threshold at each band is reached. To proceed to the next bin, the brightest pixel among the remaining pixels is selected as the starting pixel, and the same procedure is applied again.  

The above procedure is applied until no more bins can be made with the remaining pixels. In most cases, pixels around the outskirt are left without being binned. This likely caused by the insufficient number of those outskirt pixels (which typically have low S/N) left over by the previous binning process that makes binning some of them that have similar SED (within a certain $\chi^{2}$, to be described later) cannot reach the required S/N threshold. In this case, all the remaining pixels are finally binned into one bin. 

The similarity of SED shape of a pixel with index of $m$ to that of the brightest pixel with index of $b$ is evaluated with the following $\chi^{2}$ formula
%------------------------------------------------
% Equation (1): chi-square for binning
\begin{equation}
\chi^{2}=\sum_{i}\frac{(f_{m,i}-s_{mb}f_{b,i})^{2}}{\sigma_{m,i}^{2}+\sigma_{b,i}^{2}}.
\end{equation}
%-------------------------------------------------
$i$ in the above equation represents photometric band, and $f_{m,i}$ and $f_{b,i}$ are $i$-th band flux of a pixel $m$ and $b$, respectively. $\sigma_{m,i}$ and $\sigma_{b,i}$ are $i$-th band flux uncertainty of the pixel $m$ and $b$. $s_{mb}$ is a scaling factor that bring the two SEDs into a similar normalization, and it can be calculated using
%----------------------------------------- 
% Equation (2): normalization 
\begin{equation}
s_{mb}=\frac{\sum_{i}\frac{f_{m,i}f_{b,i}}{\sigma_{m,i}^{2}+\sigma_{b,i}^{2}}}{\sum_{i}\frac{f_{b,i}^{2}}{\sigma_{m,i}^{2}+\sigma_{b,i}^{2}}}.
\end{equation}
%-------------------------------------------
If $\chi^{2}$ is smaller than a certain value ($\chi^{2}_{\rm max,bin}$ which is defined by the user), the pixels $m$ and $b$ are considered to have a similar SED shape.

\begin{figure}[ht]
\centering
\includegraphics[width=0.5\textwidth]{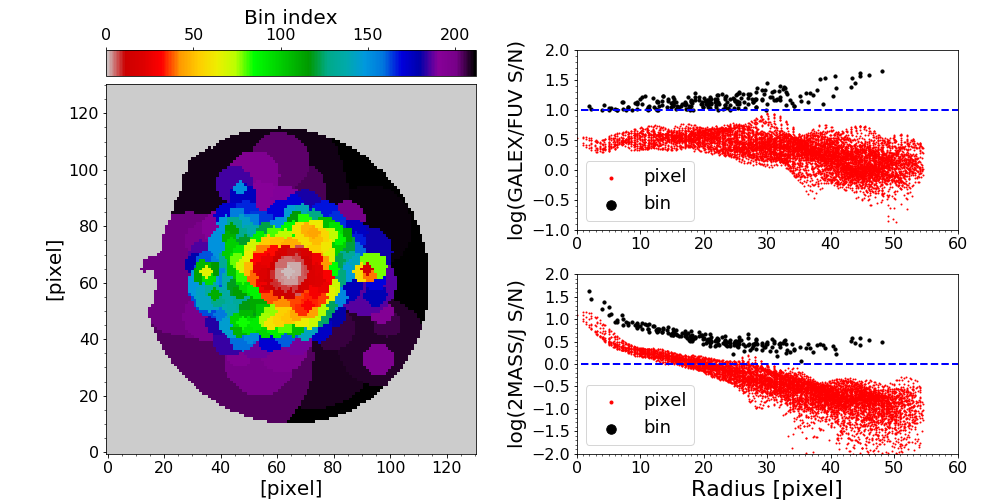}
\includegraphics[width=0.5\textwidth]{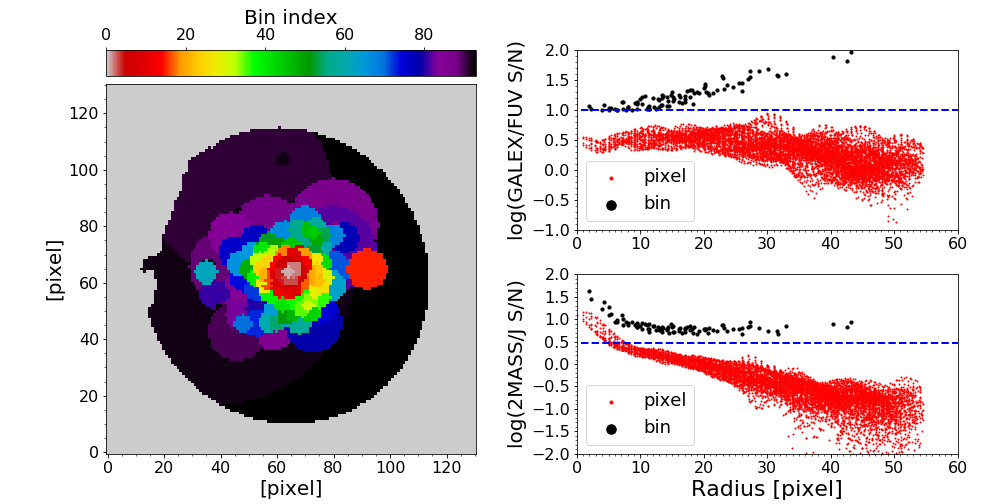}
\caption{Results of pixel binning for NGC 309 obtained with the \texttt{piXedfit\_bin} module. The top and bottom panels show results of pixel binning with requirements that only differ in the $\text{S}/\text{N}$ thresholds for the three 2MASS bands. The top (bottom) panel uses $\text{S}/\text{N}$ thresholds of $1$ ($3$) for the 2MASS bands. The two pixel binning use the same $\text{S}/\text{N}$ thresholds of $10$ for all other bands. The other requirements are the same for the two binning: $D_{\rm min,bin}$ of $4$ pixels, reduced $\chi^{2}_{\rm max,bin}$ limit of $3.3$ in the SED shape similarity check.}
\label{fig:demo_pixbin}
\end{figure}

Figure~\ref{fig:demo_pixbin} shows two pixel binning results for the NGC 309 obtained with binning requirements that only differ in $\text{S}/\text{N}$ thresholds for the three 2MASS bands. The pixel binning results in the top and bottom panels use 2MASS $\text{S}/\text{N}$ thresholds of $1$ and $3$, respectively. The $\text{S}/\text{N}$ threshold for the rest of the photometric bands is set to $10$ (see Figure~\ref{fig:maps_multiband_fluxes} for the set of the photometric bands). The other requirements are the same for the two binning: $D_{\rm min,bin}$ of $4$ pixels and reduced $\chi^{2}_{\rm max,bin}$ limit of $3.3$ in the SED shape similarity check. 

The $\text{S}/\text{N}$ ratios in the FUV and $J$ of the original pixels and bins are shown on the right side of each panel. The blue lines show $\text{S}/\text{N}$ thresholds. The pixel binning scheme is able to meet the minimum $\text{S}/\text{N}$ requirement. A general trend is that the bin size increases with radius from the galaxy's center, which can be understood because the $\text{S}/\text{N}$ of pixels decreases with radius, and thus more pixels are needed in a bin to reach the $\text{S}/\text{N}$ threshold. In this example, the 2MASS bands determine the overall result of the pixel binning because they are the shallowest (i.e.,~having lowest $\text{S}/\text{N}$) among the photometric bands used in this analysis. Due to the similarity SED shape requirement, the pixel binning map roughly reconstruct the spiral arms structure (where young stellar populations are), especially in the first binning analysis (top left panel).   

For binning a spectrophotometric data cube, we use the pixel binning map obtained with multi-band images (described above) as a reference to bin the spectrophotometric SEDs of pixels, so that the spectroscopy and photometry of a bin are consistent. For a bin in which some of the member pixels do not have spectroscopic SED, we only assign spectrophotometric SED to a bin in which at least $90\%$ of the member pixels have spectroscopic SED. The derived spatial binning map together with the fluxes and flux uncertainties are then saved into a multi-extension FITS file.  

%%=======================================%%%
\subsection{\texttt{piXedfit\_model}: Generating Model SEDs} \label{sec:piXedfit_model}
\verb|piXedfit_model| is a module designed for generating a model SED of a Composite Stellar Population (CSP) from a given set of input parameters. 

\subsubsection{Generating Rest-frame Model Spectra} \label{sec:restframe_modelSED}
For generating model spectra, the Flexible Stellar Population Synthesis (\verb|FSPS|)\footnote{\url{https://github.com/cconroy20/fsps}} package is used \citep{2009Conroy, 2010Conroy}. For interface to the Python environment, \verb|python-fsps|\footnote{\url{http://dfm.io/python-fsps/current/}} package is used \citep{2014Foreman}. The \verb|FSPS| package provides a self-consistent modeling of galaxy's SED through a careful modeling of the physical components that produce the total luminosity output of a galaxy. Those components consist of stellar emission, nebular emission, dust emission, and emission from the dusty torus heated by the AGN. We refer reader to \citet{2009Conroy}, \citet{2010Conroy}, and \citet{2017Leja, 2018Leja} for detailed description of the SED modeling within the \verb|FSPS|. For efficiency, we do not describe in detail the ingredients of the SED modeling in this paper but present the parameters in the SED modeling and fitting in Table~\ref{Tab:parameters}\footnote{A more detailed descriptions of the ingredients in the SED modeling and the parameters associated with it are available at \url{https://pixedfit.readthedocs.io/en/latest/ingredients_model.html}}.   

In generating spectra of the Simple Stellar Population (SSP), \verb|piXedfit_model| uses an option in the \texttt{FSPS} that allows interpolation of SSP spectra between the $Z$ grids available in the isochrone and spectral libraries. The nebular emission modeling uses the \texttt{CLOUDY} code \citep{1998Ferland, 2013Ferland} which was implemented in the \texttt{FSPS} by \citet{2017Byler}. For the dust attenuation modeling, \verb|piXedfit_model| allows two options: \citet{2000Calzetti} and the two-component dust model of \citet{2000Charlot}. The dust emission modeling in \texttt{FSPS} assumes the energy balance principle, where the amount of energy attenuated by the dust is equal to the amount of energy re-emitted in the infrared \citep{2008daCunha}. \texttt{FSPS} uses the \citet{2007Draine} dust emission templates to describe the shape of the infrared SED. For the modeling of emission from the dusty torus heated by the AGN, \texttt{FSPS} uses AGN templates from the \citet{2008Nenkova_a, 2008Nenkova_b} \texttt{CLUMPY} models. 

Due to the rare availability of the high spatial resolution of imaging data in the infrared, the dust emission and AGN dusty torus emission components are not applicable in most of the spatially resolved SED fitting implementation. We still include dust emission and AGN dusty torus emission in the \verb|piXedfit_model| because this module together with \verb|piXedfit_fitting| can be used for fitting an integrated SED of a galaxy, not limited to the spatially resolved SED. In case the sufficiently high spatial resolution infrared imaging data is available and the AGN component is necessary in the SED modeling, it is possible to include the AGN component to fit only the SED of the central bin of a galaxy. Using the $D_{\rm min,bin}$ parameter in the pixel binning (see Section~\ref{sec:piXedfit_bin}), the minimum diameter of a bin can be set to be similar to the PSF FWHM of the images (which is implemented in the binning result that is shown in the top left panel of Figure~\ref{fig:demo_pixbin}). Thus, the central bin always larger than the PSF size, which supposed to enclose the AGN dusty torus component in the galaxy. 

\begin{figure}[ht]
\centering
\includegraphics[width=0.5\textwidth]{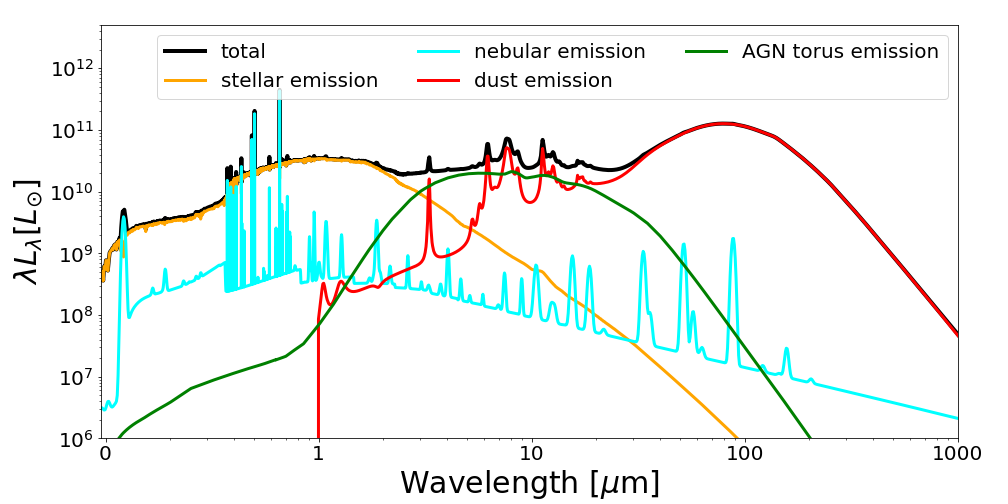}
\caption{Example of a rest-frame model spectrum (black line) generated with \texttt{piXedfit\_model}. The decomposition of the spectrum to its components is shown with different colors: orange (stellar emission), cyan (nebular emission), red (dust emission), and green (AGN dusty torus emission). The model spectrum is generated assuming delayed tau SFH with $\log(\tau[\text{Gyr}])=0.115$, $\log(\text{age}_{\rm sys}[\text{Gyr}])=0.637$, $\log(Z/Z_{\odot})=-1.208$, \citet{2000Calzetti} dust attenuation law with $\hat{\tau}_{2}=1.384$, $\log(M_{*}/M_{\odot})=10.746$, $\log(f_{\rm AGN})=-0.809$, $\log(\tau_{\rm AGN})=0.957$, $\log(Q_{\rm PAH})=0.375$, $\log(U_{\rm min})=0.953$, and $\log(\gamma_{e})=-2.035$. For the meaning of the parameters please refer to Table~\ref{Tab:parameters}.}
\label{fig:plot_restspecSED_decompose}
\end{figure}

Figure~\ref{fig:plot_restspecSED_decompose} shows an example of rest-frame model spectrum (in black color) generated using the \verb|piXedfit_model| module. The model spectrum is broken down into its components: stellar emission (orange color), nebular emission (blue color), AGN dusty torus emission (green color), and dust emission (red color). Please refer to the caption for the values of the parameters used to generate the model spectrum.
              
\subsubsection{Choices for the Star Formation History (SFH)} \label{sec:choices_SFH} 
In SED fitting, the assumed SFH is one of the fundamental components yet difficult to constraint. As a fundamental component, the assumed SFH and associated priors are very influential to the inferred physical properties of galaxies, such that the robustness of the inferred parameters is dependent on whether or not the assumed SFH is flexible enough to reflect the  true SFH of the galaxies \citep[see e.g.,][]{2009Lee,2010Maraston,2012Michalowski,2014Michalowski,2013Conroy,2017Iyer,2019Carnall,2019Leja_a,2020Lower}.

The recent developments in SED fitting enable the inference of SFH (i.e.,~SFH is not only an assumption in the fitting). There have been many attempts that try to infer SFH of galaxies using SED fitting \citep[e.g.,][]{2008Dye,2015Smith,2016Pacifici,2017Iyer, 2019Iyer, 2018Carnall, 2018Dressler, 2019Leja_a, 2019Morishita}. In terms of the SFH modeling approach, the SED fitting techniques can be classified into two main categories: parametric and non-parametric SFH. The former assumes a functional form for the SFH \citep[e.g.,][]{2014Han, 2018Carnall, 2019Boquien, 2020Zhou}, while the latter do not, instead the look-back time (i.e.,~stellar ages) is gridded and the SFR of each time grid is let free in the fitting (e.g.,~\texttt{VESPA}, \citealt{2007Tojeiro}; \citealt{2016Dressler}; \texttt{prospector}, \citealt{2017Leja}; \citealt{2018Chauke}; \texttt{gsf}, \citealt{2019Morishita}; \texttt{Dense Basis}, \citealt{2017Iyer}, \citealt{2019Iyer}), or another way is using a set of SSPs with various ages and metallicities to fit the observed SED (typically a spectrum, e.g.,~\texttt{STARLIGHT}, \citealt{2005CidFernandes}; \texttt{STECMAP}, \citealt{2006Ocvirk}; \texttt{FIREFLY}, \citealt{2017Wilkinson}). 

The parametric SFH approach has the advantage of having fewer numbers of free parameters involved in the fitting and unlimited stellar age sampling (i.e.,~time resolution in the SFH) compared to the non-parametric approach. The non-parametric approach is expected to be more flexible in reflecting the real SFH of galaxies (compared to the parametric one). However, this approach has the disadvantage of the cruder sampling of stellar ages and possibly complex degeneracies in the fitting due to large numbers of parameters involved.

Recently, \citet{2018Carnall} have shown that using the double power law SFH model can recover SFHs of simulated galaxies from the \verb|MUFASA| suite of cosmological hydrodynamical simulations. The double power law form has also been applied to fit the evolution of the cosmic SFR density \citep{2013Behroozi}. Another study by \citet{2017Diemer} showed that the log-normal SFH model can produce good fits to SFHs of simulated galaxies from the cosmological simulation Illustris. In \verb|piXedfit_model|, we adopt the parametric SFH approach, with 5 choices: exponentially declining (i.e.,~tau model), delayed tau, log-normal, Gaussian, and double power law SFHs\footnote{The functional forms of the SFH models are described in detailed at \url{https://pixedfit.readthedocs.io/en/latest/ingredients_model.html}}. The double power law SFH has the following form,
\begin{equation}  \label{eq:dpl_sfh}
SFR(t) \propto \left[ \left(\frac{t}{\tau} \right)^{\alpha} + \left(\frac{t}{\tau} \right)^{-\beta} \right]^{-1},    
\end{equation}
where $\alpha$ and $\beta$ are the falling slope, and the rising slope, respectively. The $\tau$ parameter controls the peak time. The $t$ in the above equation represent the time since the start of star formation (i.e., age of the system, $\text{age}_{\text{sys}}$).

\subsubsection{IGM Absorption, Cosmological Redshifting, and Integrating through Photometric Filters} \label{sec:cosmo_redshifting}
The rest-frame model spectra generated in the previous step are then attenuated further to account for the absorption due to the intergalactic medium (IGM) between the galaxy and the observer. The \verb|piXedfit_model| has two options for the IGM absorption: \citet{1995Madau} and \citet{2014Inoue}. The effect of cosmological redshifting and dimming is then applied to the model spectra. This will transform the spectra (that are still in unit of luminosity density, $L_{\lambda}$) into the observer frame flux density ($f_{\lambda}$). For this operation, redshift information of the galaxy is needed. However, if the redshift is unknown, it will be a free parameter in the fitting. The calculation of the luminosity distance uses the \verb|cosmology| package in the \verb|Astropy|. The last step in generating model photometric SEDs is to convolve the model spectra with the set of filter transmission functions. The current vesion of \verb|piXedfit| has a library of transmission functions for 163 photometric filters of ground-based and space-based telescopes. The user can also add a filter transmission function using a specific function in \verb|piXedfit|.

Please refer to Table~\ref{Tab:parameters} for a compilation of the parameters involved in the SED modeling and fitting.
 
\begin{table*}[ht]
\centering
\caption{Description for the parameters involved in the SED modeling and fitting}
\begin{tabular}[t]{ll}
\toprule\toprule
%\hline\hline
Parameter&Description\\
\midrule
$M_{*}$&Stellar mass\\
$Z$&Stellar metallicity\\
$t$&Evolving age ($\text{age}_{\text{sys}}$) of the stellar population\\
$\tau$&A parameter in the SFH that controls the duration of star formation\\
$T_{0}$&A parameter in the log-normal and Gaussian SFHs that controls the peak time\\
$\alpha$&A parameter in the double power law SFH that controls the slope of the falling star formation episode\\
$\beta$&A parameter in the double power law SFH that controls the slope of the rising star formation episode\\   
$\hat{\tau}_{1}$&Dust optical depth of the birth cloud in the \citet{2000Charlot} dust attenuation law\\
$\hat{\tau}_{2}$&Dust optical depth of the diffuse ISM in the \citet{2000Calzetti} and \citet{2000Charlot} dust attenuation laws\\
$n$&Power law index in the dust atttenuation curve for the diffuse ISM in the \citet{2000Charlot} dust attenuation law\\
$U$&Ionization parameter in the nebular emission modeling\\
$U_{\rm min}$&Minimum starlight intensity that illuminate the dust\\
$\gamma_{e}$&Fraction of total dust mass that is exposed to this minimum starlight intensity\\
$Q_{\rm PAH}$&Fraction of total dust mass that is in the polycyclic aromatic hydrocarbons (PAHs)\\
$f_{\rm AGN}$&AGN luminosity as a fraction of the galaxy bolometric luminosity\\
$\tau_{\rm AGN}$&Optical depth of the AGN dusty torus\\
\bottomrule
\end{tabular}
\label{Tab:parameters}
\end{table*}%

\section{SED Fitting Approach in \texttt{piXedfit}} \label{sec:SEDfit_procedure}
The SED fitting in \verb|piXedfit| is done by \verb|piXedfit_fitting| module. This module can perform SED fitting to a photometric SED as well as a spectrophotometric SED. The SED fitting approach adopted in \verb|piXedfit| is described in the following sections.  

\subsection{Bayesian Inference Method} \label{sec:bayesian_framework}
The \verb|piXedfit_fitting| module uses the Bayesian inference technique for estimating the underlying parameters of a galaxy's SED. Two important components in the Bayesian inference process are the likelihood (i.e.,~$P(X|\theta)$, which is the probability of observing the data $X$ given the model $\theta$) and prior (i.e.,~$P(\theta)$, which is the hypothesis on the probability of model $\theta$ before fitting with the data). In SED fitting, the likelihood is commonly given by the Gaussian function because of the assumption of a Gaussian form of noise. The Gaussian likelihood form is used by the majority of Bayesian SED fitting implementation, e.g.,~\citet{2003Kauffmann}, \texttt{MAGPHYS} \citep{2008daCunha}, \texttt{BayeSED} \citep{2014Han}, \texttt{BAGPIPES} \citep{2018Carnall}, \texttt{CIGALE} \citep{2005Burgarella,2009Noll,2019Boquien}.  

In \citet{2017Abdurrouf}, we implemented a different likelihood function that make use of the Student's t function. The new likelihood function has been shown to be able to give a better recovery of the SFR in the fitting tests using mock SEDs and better matching to the SFR derived from the \textit{Spitzer}/MIPS $24\mu$m flux (see Appendix A of \citealt{2017Abdurrouf}). Motivated by this result, we implement two kinds of likelihood functions in \texttt{piXedfit}: (1) Gaussian function as mentioned above, and (2) Student's t function which has the following form
\begin{equation}\label{eq:stdt_likelihood}
P(X|\theta) = \prod_{i=1}^{n} \frac{\Gamma\left(\frac{\nu+1}{2} \right)}{\sqrt{\nu \pi}\Gamma\left(\frac{\nu}{2} \right)} \left(1+\frac{\chi_{i}^{2}}{\nu}  \right)^{-\frac{\nu+1}{2}},
\end{equation}
with $\chi_{i}$ is given by
\begin{equation}
\chi_{i} = \frac{f_{X,i}-sf_{\theta,i}}{\sigma_{X,i}}.
\end{equation}
The $n$ represents number of bands (in case of photometric SED) or wavelength points (photometric bands and wavelength grids of the spectrum, in case of spectrophotometric SED), while $f_{X,i}$ and $\sigma_{X,i}$ represent the observed flux and its associated uncertainty in a given band or wavelength $i$, respectively. In case of fitting to a spectrum (or spectrophotometric SED), only the spectral continuum (or spectral continuum and photometric SED) is fitted. A certain window (default of $\pm 10\text{\normalfont\AA}$) around all possible emission lines (based on the list of emission lines wavelengths from the \texttt{FSPS}) is used to exclude emission lines in the fitting. The $f_{\theta,i}$ and $s$ are flux of model SED in band or wavelength point $i$ and a scaling factor that bring the model SED in overall similar normalization as that of the observed SED, respectively. Since model SED generated with \texttt{FSPS} is normalized to $1M_{\odot}$, so $s$ corresponds to the stellar mass.

The $\nu$ represents the degree of freedom which should be specified by the user. A large value of $\nu$ will give a likelihood function similar to that of Gaussian, while a small value of $\nu$ will give heavier tails in the likelihood distribution (compared to the Gaussian one). In Appendix~\ref{apdx:find_dof_stdt}, we compare performances of various fitting approaches and determine the best value for $\nu$. We find that $\nu\sim 1-3$ give overall robust and stable inference of parameters.

The flux uncertainty ($\sigma_{X,i}$) is not just taken from the observational error, which is often an underestimation, but also consider the systematic uncertainties which come from the observational procedure (e.g.,~associated with image processing) and the SED modeling procedures. We assume that the bulk of the systematic uncertainties is a multiplicative factor of the observed fluxes such that $\sigma_{sys,i}=\text{err}_{sys} \times \sigma_{X,i}$, following \citet{2019Han}. We do not set the $\text{err}_{sys}$ as a free paremeter in the fitting considering that it can possibly add a degeneracy in the fitting process, instead we fixed it to a certain value that is obtained from a fitting test that can be done either to each individual galaxies or to one galaxy representative of a whole sample. Practically, in the fitting test we vary the $\text{err}_{sys}$ such that the reduced $\chi^{2}$ of the best-fit model SED is below $\sim 2.0$. Without adding such systematic uncertainties, it is quite often to find cases where the reduced $\chi^{2}$ of the best-fit model SED is large while the fluxes residuals are actually very small. From analysis of 20 local galaxies (to be presented in Section~\ref{sec:test_with_specphoto_data}), we find that $\text{err}_{sys} \lesssim 0.15$ is enough to reach the required reduced $\chi^{2}$ mentioned above.  

In the default setting and in the analysis throughout this paper, a flat prior over a certain range is assumed for each parameter. For versatility, \verb|piXedfit_fitting| can also adapt with the priors given by the user in array or a text file format.  

\subsection{Posterior Sampling Method} \label{sec:posterior_sampling}
The main task in Bayesian parameter inference is to solve for the posterior probability distribution function of each parameter. Commonly, a sampling method is used to reconstruct the posteriors. In the SED fitting application, there are at least three approaches adopted for the posterior sampling: the gridding method \citep[e.g.,][]{2019Boquien,2020Chen}, MCMC \citep[e.g.,][]{2011Acquaviva,2017Leja,2019Morishita}, and nested sampling \citep[e.g.,][]{2014Han, 2018Carnall, 2019Leja}. 

In the gridding method, each parameter space is divided into a number of grids, then model SEDs are generated for all the possible combinations of the parameters grids. One of the advantages of the gridding method is that it could fit a large number of SEDs quickly, especially if the set of model SEDs (with many redshift grids) are generated before the fitting. The disadvantage of this method is that it typically requires a large number of parameters grids (and so the number of model SEDs) in order for the sampling to be complete, especially for high dimensional parameter space. In the MCMC fitting, the $N$ dimensional parameters are explored by random walks of sampler chains. Over time, the frequency of visited locations can in principle be a representative of the posterior probability function. The disadvantage of this method is that it is computationally expensive and typically slow.          

In \verb|piXedfit_fitting|, we adopt two different posterior sampling methods: MCMC and random densely-sampling of parameter space (hereafter RDSPS). Each of those methods is described in the following.

\subsubsection{Fitting with MCMC} \label{sec:mcmc_sampling}
For the MCMC sampling, we use \verb|emcee|\footnote{\url{https://github.com/dfm/emcee}} package by \citet{2013Foreman, 2018Foreman_zenodo, 2019Foreman}. Before running the MCMC sampling, an initial fitting is done using the $\chi^{2}$ minimization technique to get an initial guess and set initial positions for the MCMC walkers. For this fitting, a set of pre-calculated model SEDs (to be described in Section~\ref{sec:random_uniform}) is used. 
The initial positions for the MCMC walkers are defined by a small asymmetric Gaussian ``ball'' with a $\sigma=0.08\times W$ around the best-fit parameters obtained from the initial fitting. The $W$ is the width (i.e.,~prior range) of a parameter space.          

The next step is running the MCMC. The number of MCMC walkers and steps should be defined by the user. When the MCMC is running, a model likelihood has to be supplied for each ensemble of $N$ parameter values that are generated. In this case, we use the Gaussian likelihood function for calculating the model likelihood. The MCMC sampling will finish when every walker has completed the specified number of steps. The results of MCMC sampling is the sampler chains which record the locations in the parameter space that are visited by the walkers throughout the process. From these sampler chains, the posterior probability distribution of each parameter can be constructed. The inferred value for each parameter is then obtined from the median of the posterior, while the uncertainty is defined by the range given by the 16th and 84th percentiles. In order to make the calculation efficient, the parallelization module in \verb|emcee| is implemented. 

\subsubsection{Random Densely-sampling of Parameter Space (RDSPS)} \label{sec:random_uniform}
The second sampling method we adopt is the RDSPS method, which is a simple sampling method inspired by the gridding method described previously. Unlike the gridding method which defines fixed grids of values for each parameter, the RDSPS method draws random values uniformly within the prior range in each parameter. For generating $N_{\rm mod}$ number of model SEDs with $N$ number of parameters, an $N_{\rm mod}$ number of random values are generated for each parameter. Then, those $N$ arrays of parameters are randomly connected with each other to construct the library of model SEDs. The reason of using the RDSPS method over the gridding method is its efficiency. With a smaller number of generated models (e.g.,~$\sim$500000 for 9 free parameters), sub regions in each parameter axis can be represented by at least several models. 

In order to reduce the computation time, large number of model SEDs are calculated and stored into FITS files. The models are calculated in many grid of redshifts with increment of $0.002$. A set of model SEDs with the same redshift is stored into one FITS file. Then, this library of model SEDs can be used for fitting all the galaxies in a sample. In the fitting where redshift of the galaxy is known, model SEDs are calculated for that redshift by applying cubic spline interpolation from the set of pre-calculated model SEDs. Otherwise, the redshift will be set as a free parameter. In the spatially resolved SED fitting application, for higher accuracy, it is also possible to generate a set of model SEDs for each galaxy based on the known redshift of the galaxy. Then this set of model SEDs is used for fitting all the spatial bins of the galaxy.

The next step in the fitting is to calculate the posterior probability of each model. For fitting with the RDSPS method, we allow two kinds of likelihood functions: Gaussian and the Student's t functions. In the calculation of model likelihood, the normalization ($s$) of a model SED is calculated from the analytical solution for minimizing the $\chi^{2}$ (see e.g., Eq. 7 in \citealt{2012Sawicki}). We do not set $s$ as free in the fitting for the sake of efficiency.

After calculating the posterior probability of each model, the inferred value of each parameter is obtained from weighted averaging with model posterior serving as the weight for the model. The uncertainty is estimated from the weighted standard deviation. For fast fitting performance, we have incorporated the parallel processing module, namely message passing interface (MPI) in this SED fitting module.

\subsection{\texttt{piXedfit\_analysis}: Visualization of Fitting Result} \label{sec:piXedfit_analysis}
The output of the fitting process with the \verb|piXedfit_fitting| module is a FITS file containing sampler chains (in the case of fitting with MCMC) or posterior probabilities of model SEDs (in the case of fitting with the RDSPS method). The FITS file can then be used for further analysis, such as deriving inferred values of parameters and visualization of the fitting, the latter task can be done with \verb|piXedfit_analysis| module.

For visualizing the fitting results with MCMC, 3 kinds of plots can be made using the \verb|piXedfit_analysis| module: corner plot, SED plot, and SFH plot. The corner plot shows the posterior probability distributions (constructed from the sampler chains) of individual parameters (as 1D histograms) as well as joint posterior probability distributions of every pair of two parameters (in 2D). In the corner plot, inferred values of parameters (from median of the posteriors), the uncertainty (16th--84th percentiles of the posteriors) are shown with black vertical line and gray shaded area in the 1D histograms, respectively. For producing the SED plot, an ensemble of $200$ sampler chains is randomly picked from the full MCMC sampler chains, then their spectra are generated. The median posterior model SED (spectrum as well as photometric SED) and its uncertainty are then obtained by taking median, 16th and 84th percentiles from the ensemble of spectra. The residual, which is $(f_{X,i}-f_{\theta,i})/f_{X,i}$, is also shown in the SED plot (see Section~\ref{sec:bayesian_framework} for the definitions of $f_{X,i}$ and $f_{\theta,i}$). For producing the SFH plot, the inferred SFH is derived by first randomly picking $200$ sampler chains from the full MCMC sampler chains, then the SFHs associated with the sampler chains are calculated. The median, 16th and 84th percentiles are then calculated from the ensemble of SFHs at each time step. The median is then used as the inferred SFH, while the area between the 16th and 84th percentiles is used as the associated uncertainty. For fitting with the RDSPS method, currently, only the SED plot can be produced in which the best-fit model SED is obtained from the model with lowest $\chi^{2}$. Example of the corner plot, SED plot, and SFH plot can be seen in Figures~\ref{fig:demo_pixedfit_tng} and~\ref{fig:demo_pixedfit_ngc309}.

\section{Testing the SED Fitting Performance Using Mock SEDs of IllustrisTNG Galaxies} \label{sec:tetsfit_illustris_TNG}
In this section, we use FUV--NIR mock SEDs of the IllustrisTNG (hereafter TNG) galaxies to test the performance of the \verb|piXedfit_fitting| module in terms of its abilities in parameter inference and SFH reconstruction. We leave the fitting experiment that uses mock FUV--FIR SEDs for future work. 

\subsection{Generating Mock SEDs of TNG Galaxies} \label{sec:generate_mockSEDs_TNG}
The IllustrisTNG simulations\footnote{\url{http://www.tng-project.org}}\citep{2018Marinacci, 2018Naiman, 2018Nelson, 2018Pillepich, 2018Springel, 2019Nelson} are a suite of cosmological hydrodynamical simulations that model a range of physical processes involved in the formation of galaxies. In order to test the performance of the SED fitting using \verb|piXedfit_fitting| in inferring the galaxy properties, we generate mock SED of TNG galaxies and then fit them with the \verb|piXedfit_fitting| module to see whether the inferred parameters can recover the true properties of the TNG galaxies. Furthermore, having realistic SFH from the TNG galaxies, we can also test the performance of the \verb|piXedfit_fitting| module in reconstructing the SFH of a galaxy. 

For this fitting test, we use the fiducial TNG100 simulation, which has a volume of $\sim100^3$ comoving Mpc and a baryon mass resolution of $1.4\times10^6 M_\odot$. We select 300 galaxies from the TNG100 simulation. More specifically, we select 100, 80, 60, and 40 galaxies in every $0.5$ dex bin in $M_{*}$ between $10^{9}$ and $10^{11} M_{\odot}$ and other 20 galaxies more massive than $10^{11.5}M_{\odot}$. The number is somewhat arbitrary, simply to reflect that there are more low-mass galaxies than high-mass ones. In each mass bin, we first rank all TNG galaxies by their sSFR and choose target number of galaxies equally spacing in terms of the percentiles in sSFR. In this way, the selected galaxies cover the entire sSFR range.   

The mock spectra of TNG galaxies are created by regarding a stellar particle as an SSP, then generating the spectrum of each stellar particle using FSPS. In generating the SSP spectra, Padova isochrones \citep{2000Girardi, 2007Marigo, 2008Marigo}, MILES stellar spectral library \citep{2006Sanchez-Blazquez,2011Falcon}, and \citet{2003Chabrier} IMF are assumed. The integrated spectrum of a galaxy is then obtained by summing up the spectra of gravitationally-bound stellar particles in a subhalo associated with the galaxy. We assume a redshift of $0.001$. To mimic the dust attenuation effect, we assign each galaxy with a random value of dust optical depth ($\hat{\tau}_{2}$) and then apply the \citet{2000Calzetti} dust attenuation law to the galaxy's spectrum. The random values of $\hat{\tau}_{2}$ are uniformly distributed between $0$ and $2.5$.

For this fitting test, we generate two kinds of mock SEDs: photometric and spectrophotometric SEDs. The photometric SEDs are obtained by convolving the synthetic spectra with 12 broad-band filters: GALEX (FUV, NUV),  SDSS ($u$, $g$, $r$, $i$, $z$), 2MASS ($J$, $H$, $K_{s}$), and WISE ($W1$, $W2$). For the spectrophotometric SED, the photometric SED is created with the above procedure, while the mock spectrum is created to match the characteristic of MaNGA spectrum. To mimic the observational noise, a Gaussian noise is injected to the SEDs (both photometric and spectroscopic) by randomly perturbing each flux point from the original value by drawing from a Gaussian distribution with standard deviation dictated by the flux uncertainty. We assign each SED (either photometric or spectroscopic) with a $\text{S}/\text{N}$ ratio of $10$. We create the mock FUV--NIR SEDs with the similar setting as that provided in the \verb|piXedfit_model| because we only focus on testing the performance of the fittting algorithm of \texttt{piXedfit\_fitting} module.

\subsection{SED Fitting Analysis of TNG Galaxies} \label{sec:SEDfit_analysis_TNG}
We fit the synthetic SEDs with the \verb|piXedfit_fitting| module using the same assumptions of the IMF, spectral library, isochrones, and dust attenuation law as those used for creating the synthetic SEDs. For the SFH, we use the double power law model. We choose double power law SFH form because of its flexibility in the rising and falling phases. Since the wavelength of the mock SEDs ranges from FUV to NIR, we turn off the AGN dusty torus emission and the dust emission modeling in the fitting. This leaves us with seven free parameters: $Z$, $\tau$, $t$ ($\text{age}_{\text{sys}}$), $\hat{\tau}_{2}$, $\alpha$, $\beta$, and $M_{*}$. Flat priors within a given range is assumed for all the parameters. Logarithmic sampling is applied to all the parameters, except for $\hat{\tau}_{2}$. The assumed parameters ranges for the priors are as follows: $\log(Z/Z_{\odot})=[-0.5,0.42]$, $\log(\tau)=[-3.0,0.6]$, $\log(t)=[0.9,1.14]$, $\hat{\tau}_{2}=[0.0,3.0]$, $\log(\alpha)=[-2.0,1.0]$, and $\log(\beta)=[-2.0,1.0]$. For the $M_{*}$, we use a flat prior in logarithmic scale within a range of $\log(M_{*})=[\log(s_{\text{best}})-2,\log(s_{\text{best}})+2]$, with $s_{\text{best}}$ is the normalization obtained from the initial fitting with the $\chi^{2}$ minimization technique (see Section~\ref{sec:mcmc_sampling}). 

In order to compare the performances of various fitting approaches provided within \verb|piXedfit_fitting|, we do the SED fitting with 8 different fitting approaches. These cover the two posterior sampling methods (MCMC and RDSPS), the two likelihood functions (Gaussian and Student's t) in the RDSPS method, and 6 values of degree of freedoms ($\nu$) for the Student's t likelihood function: $0.3$, $1.0$, $2.0$, $3.0$, $5.0$, and $10.0$.  In the MCMC fitting, the number of walkers and steps are 100 and 1000, respectively.           

\begin{figure*}[ht]
\centering
\includegraphics[width=1.0\textwidth]{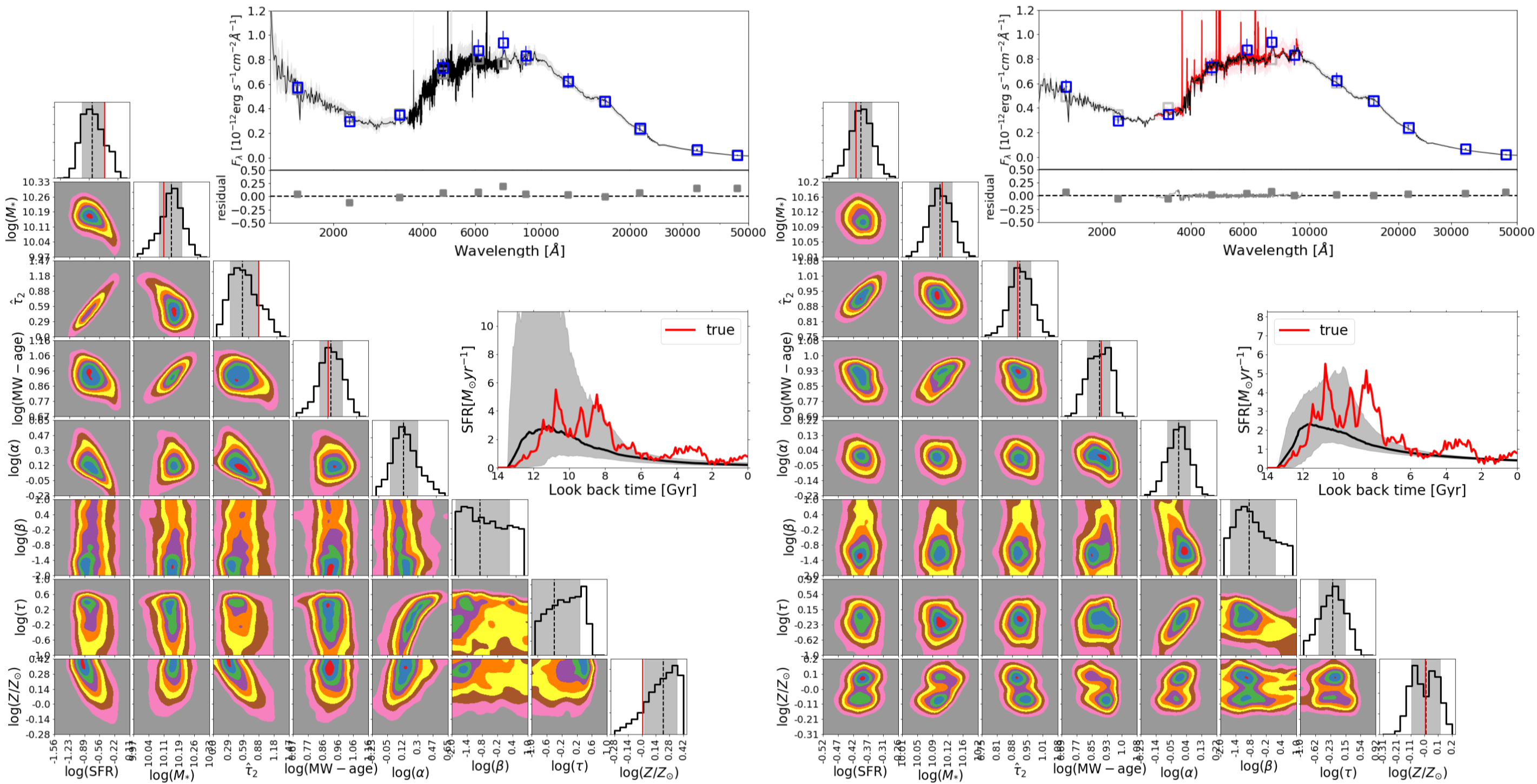}
\caption{Example of fitting results to mock photometric SED (left) and spectrophotometric SED (right) of a TNG galaxy. The fitting uses MCMC method. In each fitting result, three kinds of plots are shown: corner plot, SED plot, and SFH plot. Please see the text for description of the symbols in each plot.}
\label{fig:demo_pixedfit_tng}
\end{figure*}

Figure~\ref{fig:demo_pixedfit_tng} shows fitting results to a mock photometric SED (left) and spectrophotometric SED (right) of a TNG galaxy. The fitting uses MCMC technique. In each side, three plots are shown: a corner plot, an SED plot, and an SFH plot. The three plots are made using the \verb|piXedfit_analysis| module (see Section~\ref{sec:piXedfit_analysis}). In the corner plot, the black vertical dashed lines and the shaded area represent median and 16th--84th percentiles. The red vertical lines show the true values. The SED plot shows the mock photometric SED (blue squares), the mock spectrum (red line, in case of the right panel), the median posterior model spectrum (black line), and the median posterior model photometric SED (gray squares). The residual is given by $(f_{X,i}-f_{\theta,i})/f_{X,i}$ (see Section~\ref{sec:piXedfit_analysis}). In the SFH plot, the black line and gray shaded area represent the inferred SFH and its uncertainty, while the red line shows the true SFH. The figure shows that the SED fitting with the \verb|piXedfit_fitting| module can recover the true properties and the overall trend of the SFH of the TNG galaxy well. The addition of the synthetic spectrum in the so-called spectrophotometric SED can add more constraining power in the fitting process and result in better contraints of the $Z$ and SFH parameters (i.e.,~$\alpha$, $\beta$, and $\tau$). The fitting with spectrophotometric SED can also reveal a bimodality in the posterior probability distribution of metallicity.                

\subsection{Recovering Physical Properties of the TNG Galaxies} \label{sec:recover_params_TNG}
For the first test, in this section we compare the inferred parameters obtained from fitting and the true properties of the TNG galaxies. 
The metallicity and age of a TNG galaxy are obtained by mass-weighted averaging over the metallicities and ages of the stellar particles, respectively. The SFR of a TNG galaxy is estimated from the amount of stellar mass formed over the last $50$ Myr, based on the formation times of individual stellar particles.      

\begin{figure*}[ht]
\centering
\includegraphics[width=0.9\textwidth]{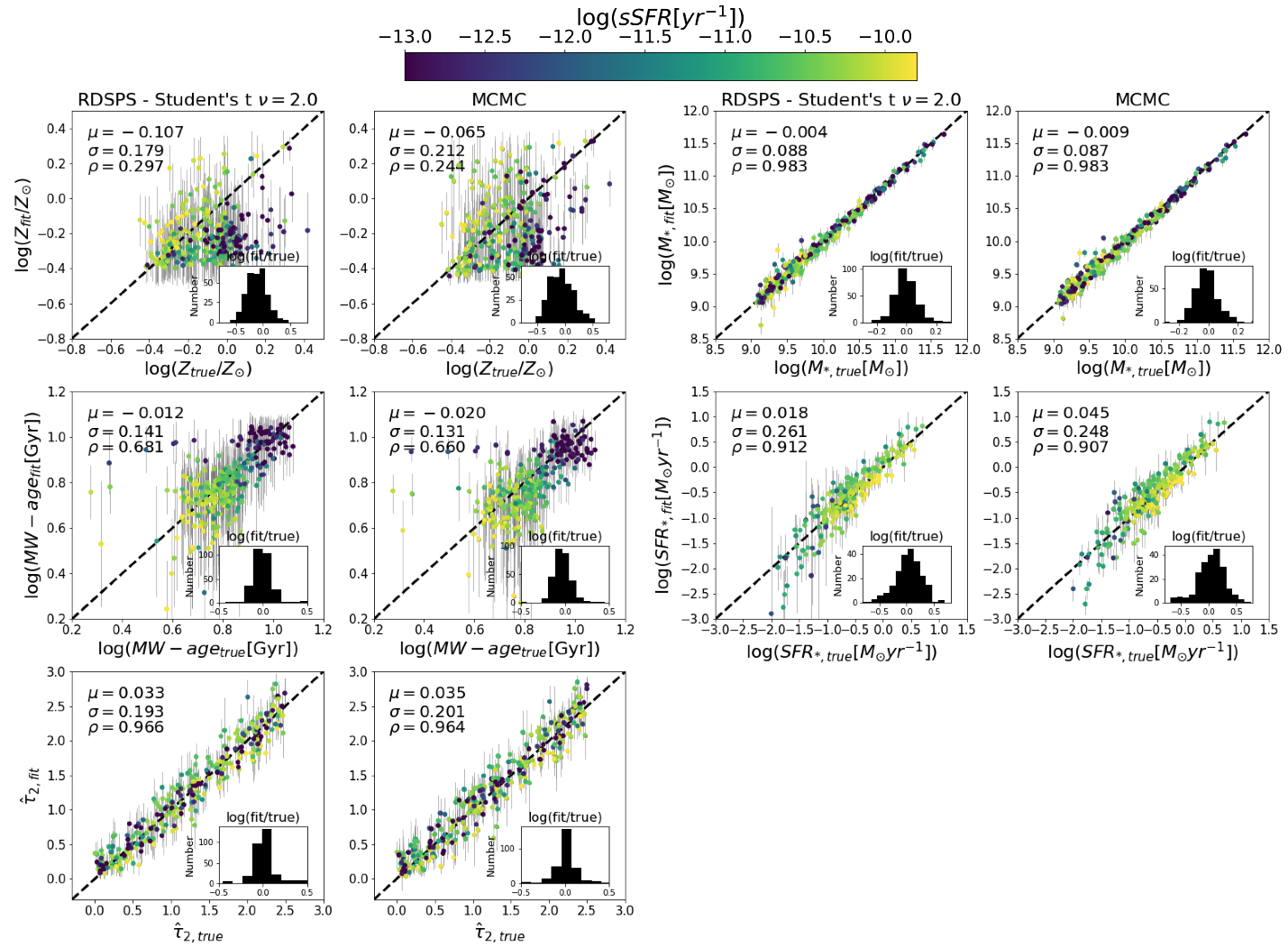}
\caption{Comparisons between the inferred parameters from fitting and the true properties of the TNG galaxies. Results from fitting with two approaches (RDSPS that uses Student's t likelihood with $\nu=2.0$ and the MCMC) are shown in this figure. The color-coding is based on the sSFR of the TNG galaxies. In each panel, a histogram of the logarithmic ratio between the inferred values from fitting and the true values, and its parameters ($\mu$, $\sigma$, $\rho$) are shown.}
\label{fig:plt_tfit_TNG_coll}
\end{figure*}

Figure~\ref{fig:plt_tfit_TNG_coll} shows direct comparisons between the inferred parameters obtained from fitting to mock photometric SEDs and the true values for two fitting approcahes: the RDSPS that uses Student's t likelihood with $\nu=2.0$ (first and third columns) and the MCMC (second and fourth columns). We discuss the fitting results obtained with the other 6 fitting approaches in Appendix~\ref{apdx:find_dof_stdt}. In brief, we found that all the 8 fitting approaches can recover the true properties of the TNG galaxies well. Though, we find an indication that the fitting approach of RDSPS that uses the Student's t likelihood function with $\nu \sim 1-3$ can outperform the RDSPS that use other likelihood functions and give results that are broadly consistent with the MCMC method, but with much faster performance.        

In Figure~\ref{fig:plt_tfit_TNG_coll}, the scattered data are color-coded based on the sSFR of the TNG galaxies. The inferred mass-weighted age is calculated by weighting age of stars (i.e.,~look back times in the SFH) with their stellar masses at the birth time (i.e.,~amount of stellar masses produced at the look back time). To assess the goodness of the parameter recovery, we calculate mean offset ($\mu$), scatter (i.e.,~standard deviation, $\sigma$), and the Spearman rank-order correlation coefficient ($\rho$, which is calculated using the \texttt{SciPy} package, \citealt{2020Virtanen}). The coefficient $\rho$ is a nonparametric measure of the monotonicity of the relationship between two datasets. The histogram for the logarithmic ratio and the associated $\mu$, $\sigma$, and $\rho$ values are shown along with the scattered data.       

The figure shows that, overall, the inferred parameters by the fittting to the photometric SEDs with the two approaches can recover the true values quite well, except $Z$ in which the true values are only broadly followed by the inferred values from fitting, though with small median offset ($\sim 0.1$ dex) and scatter ($\sim 0.2$ dex). Among the parameters, the $M_{*}$ is the best recovered, corroborated by the small offset (absolute value of $\mu$ of $\lesssim 0.01$ dex), small scatter ($\sim 0.09$ dex), and high value (close to unity) of $\rho$ ($\sim 0.98$). The $\hat{\tau}_{2}$ and SFR are also successfully recovered by the fitting. While overall the mass-weighted age is well recovered, there is a trend of increasing scatter toward the galaxies with young stellar populations. The color-coding indicates that there is a notable relation between the relatively larger scatter around the low mass-weighted age region with the increase of sSFR. The outshining effect by young stars (which is abundant in galaxies with high sSFR) may be playing a role in this. In a high sSFR galaxy, which tend to have stellar population dominated by young stars, the light from the young bright stars dominates the light from the older ones, making it relatively easy to ``hide'' old stellar populations and consequently it is more difficult to infer SFH of the galaxy \citep[see e.g.,~][]{1998Sawicki,2001Papovich,2010Maraston,2013Conroy}. However, it is also revealed that the $Z$ inference tend to be better for the high sSFR galaxies than that for the low sSFR ones.

The difficulty in inferring metallicities from SED fitting with photometric SED alone has also been reported in the literature, e.g., \citet[][their Fig. 11]{2012Pacifici}, who fit mock optical photometry with a set of model SEDs of galaxies that are drawn from a semi analytical model, which exhibit complex SFHs, \citet[][their Fig. 18]{2014Han}, who fit mock FUV--NIR SEDs with \texttt{BayeSED} that uses the nested sampling method, and \citet[][their Fig. 11]{2018Smith}, who employed \texttt{MAGPHYS} to perform pixel-by-pixel SED fitting to a set of FUV--FIR synthetic images constructed from a zoom-in simulation of an isolated disk galaxy. Despite the wide wavelength coverage (which can be expected to break the well-known age--metallicity--dust attenuation degeneracy) implemented in \citet{2018Smith}, the inferred metallicity from SED fitting is systematically underestimated compared to the true values. 

\citet{2014Michalowski} evaluated the $M_{*}$ inference of various SED fitting codes, which have various assumed SFH models, using the synthetic FUV--FIR photometric SEDs of simulated galaxies. The median offsets and scatters in the $M_{*}$ comparisons have ranges of $0.01-0.2$ dex and $0.09-0.4$ dex, respectively. \citet{2020Lower} applied non-parametric SFH model with \texttt{prospector} to fit the synthetic FUV--FIR photometric SEDs of simulated galaxies and obtained median offset of $0.02$ dex and scatter of $0.13$ dex. Our $M_{*}$ inference has a smaller offset and scatter than that obtained in the above studies. Though, the more comprehensive (i.e.,~realistic) simulation of dust component (through the radiative transfer technique) in the construction of mock SEDs that is implemented in the above studies might add more complexity in the fitting test.        

\begin{figure*}[ht]
\centering
\includegraphics[width=0.9\textwidth]{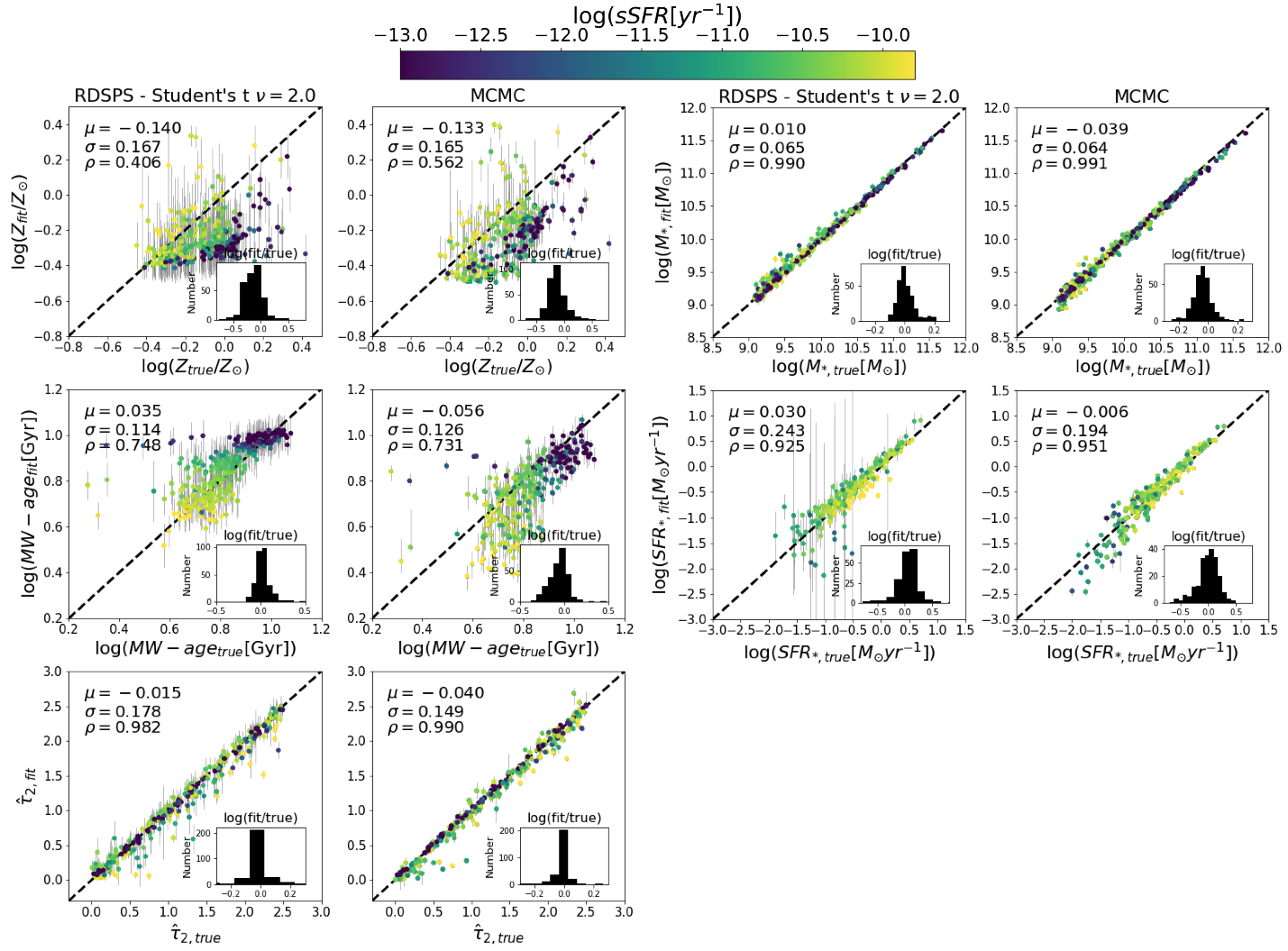}
\caption{Similar to Figure~\ref{fig:plt_tfit_TNG_coll}, but now the fitting is done to the mock spectrophotometric SEDs of the TNG galaxies.}
\label{fig:plt_specphoto_tfit_TNG_coll}
\end{figure*}

In order to investigate the effect of the inclusion of spectrum into the SED on the performance of the parameters inference, we do the same fitting tests to the mock spectrophotometric SEDs of the TNG galaxies. In fitting a spectrophotometric SED of a galaxy, only spectral continuum is fitted simultaneously with the photometric SED (see Section~\ref{sec:bayesian_framework}). Figure~\ref{fig:plt_specphoto_tfit_TNG_coll} shows the comparison between the inferred parameters obtained from the fitting and the true values. The format of this figure is the same as that of Figure~\ref{fig:plt_tfit_TNG_coll}. Overall, we see improvements on the inference of all the parameters (with the two fitting approaches) over what is obtained with the photometric data only, corroborated by the smaller scatter and higher $\rho$ value, though slightly higher offset. The significant increase in $\rho$ value of $Z$ suggests that the inferred $Z$ become better inline with the true $Z$. This happens at the same time with the decreasing scatter of $\hat{\tau}_{2}$, which suggests that the inclusion of spectrum can potentially break the degeneracies in the fitting process. This result agrees with \citet{2012Pacifici} who found that SED fitting using mock optical spectroscopy significantly improve the parameters inference over the one that only use photometry.

\subsection{Recovering SFHs of the IllustrisTNG Galaxies} \label{sec:recover_SFH_TNG}
\begin{figure*}[ht]
\centering
\includegraphics[width=1.0\textwidth]{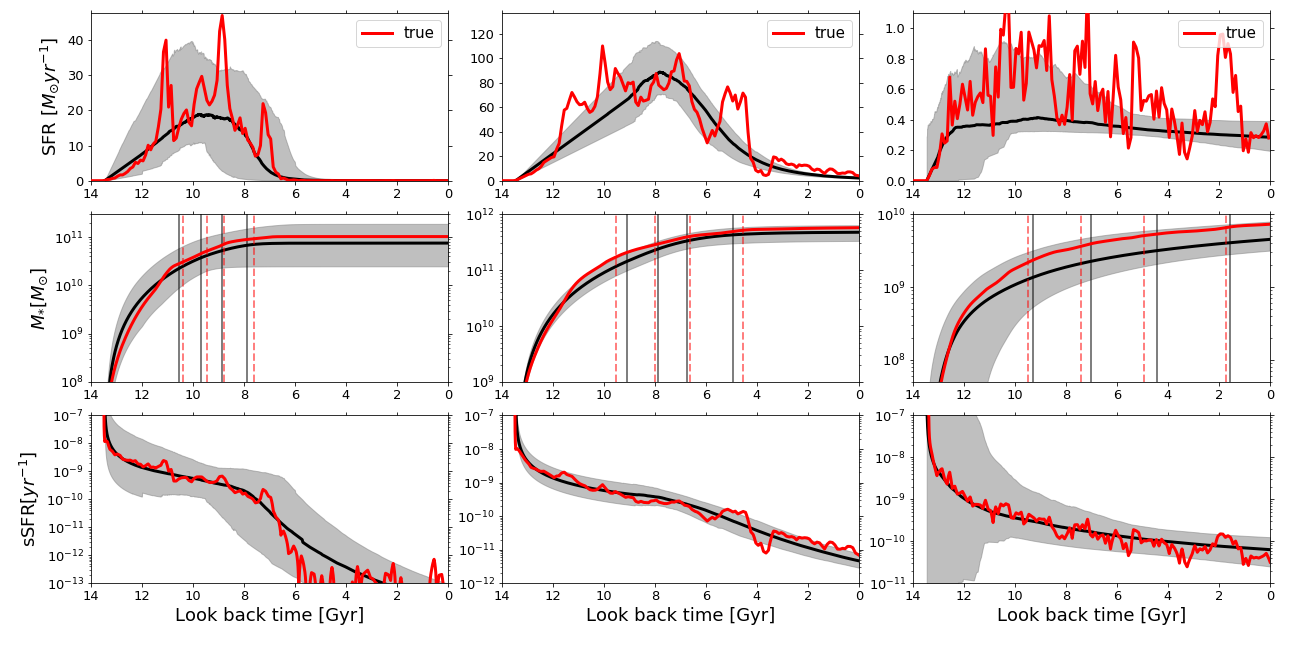}
\caption{Comparison of the inferred SFH (first row), $M_{*}(t)$ (second row), and $\text{sSFR}(t)$ (third row) obtained from fitting to the spectrophotometric SEDs of 3 TNG galaxies using the \texttt{piXedfit\_fitting} module and the true histories. The vertical red dashed lines and black lines are the true and inferred look-back times when the galaxies were still having $M_{*}$ of $30\%$, $50\%$, $70\%$, and $90\%$ of the current $M_{*}$.}
\label{fig:demo_recov_SFH}
\end{figure*}

In this section, we test the performance of the \verb|piXedfit_fitting| module in terms of its ability of inferring the SFH of galaxies. The way we do this is by comparing the inferred SFH with the true SFH of the TNG galaxies. Figure~\ref{fig:demo_recov_SFH} shows examples of SFHs (black lines and gray shaded areas in the first row) inferred by the MCMC fitting using the \verb|piXedfit_fitting| module to spectrophotometric SEDs of three TNG galaxies. In each panel in the figure, the black line represents the median, while the gray shaded area represents the uncertainty. The true SFHs of the TNG galaxies are shown by the red lines. The SFH of TNG galaxy is calculated with time steps of $100$ Myr. In the second and third rows, the histories of the stellar mass growth ($M_{*}(t)$) and sSFR ($\text{sSFR}(t)$) are shown, respectively. They are derived from the inferred SFHs. Same as in the first row, the red and black lines here represent the true and inferred histories, respectively. The vertical red dashed lines in the $M_{*}(t)$ plots are the true look-back times when the galaxies were still having $M_{*}$ of $30\%$ ($lbt_{30\%M}$), $50\%$ ($lbt_{50\%M}$), $70\%$ ($lbt_{70\%M}$), and $90\%$ ($lbt_{90\%M}$) of the current $M_{*}$, while the vertical black lines are the values inferred from the median $M_{*}(t)$. The figure shows that the inferred SFH, $M_{*}(t)$, and $\text{sSFR}(t)$ can recover the overall shape (i.e.,~the rising and falling phases) of the true histories of these three TNG galaxies well.

\begin{figure*}[ht]
\centering
\includegraphics[width=0.9\textwidth]{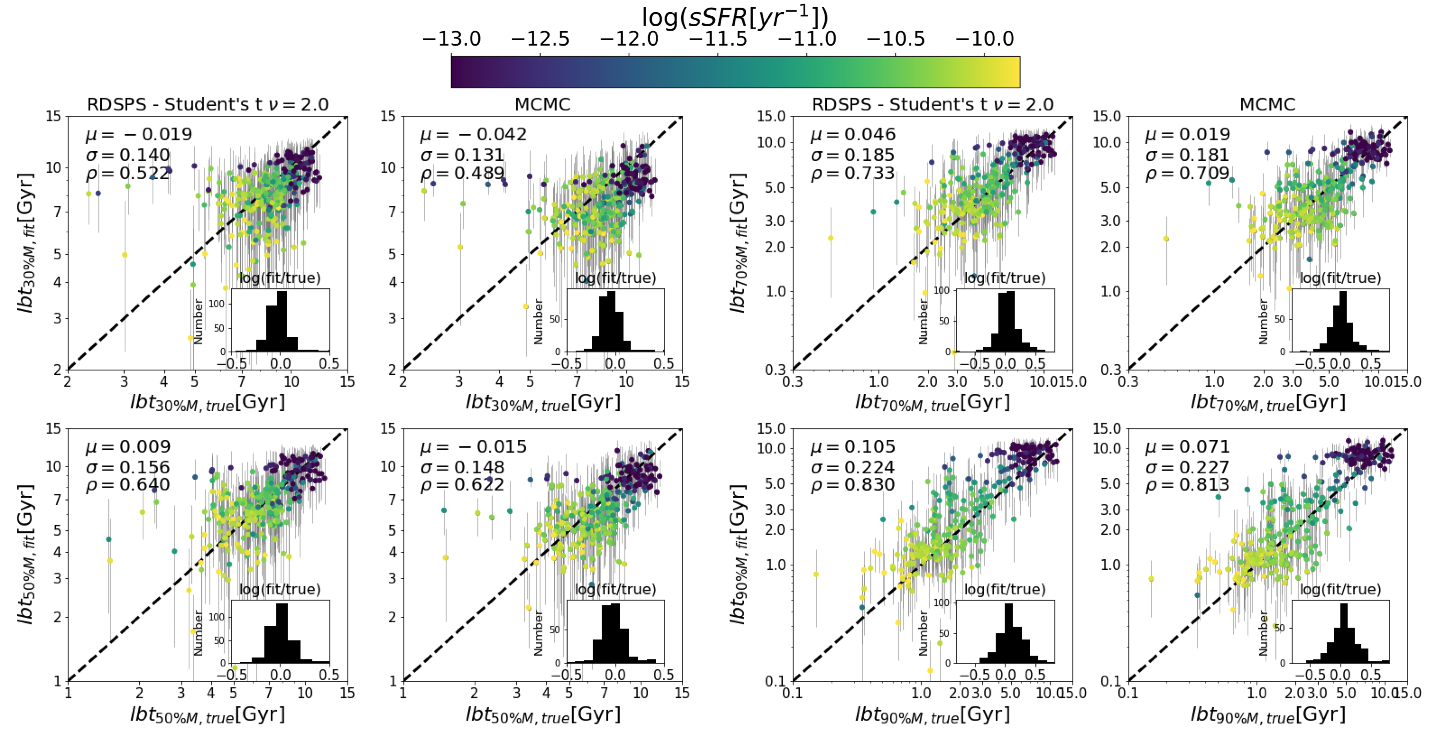}
\caption{Comparison of the inferred $lbt_{30\%M}$, $lbt_{50\%M}$, $lbt_{70\%M}$, and $lbt_{90\%M}$ from fitting to the photometric SEDs of the TNG galaxies using the \texttt{piXedfit\_fitting} module and the true values. Two fitting approaches are used: the RDSPS that uses Student's t likelihood with $\nu=2.0$ and the MCMC. The scattered data are color-coded based on the sSFR of the TNG galaxies.}
\label{fig:recov_SFH_lbt_photo}
\end{figure*}

In order to quantitatively assess the performance of the \verb|piXedfit_fitting| module in inferring the SFH, we compare the inferred and true values of the $lbt_{30\%M}$, $lbt_{50\%M}$, $lbt_{70\%M}$, and $lbt_{90\%M}$. Results from the fitting with the photometric SEDs are shown in Figure~\ref{fig:recov_SFH_lbt_photo}. This figure shows that overall, the true look-back time episodes in the $M_{*}(t)$ can be recovered well using the \verb|piXedfit_fitting| module with the two fitting approaches. The earlier look-back time episodes seem to be more difficult to recovered compared to the later ones, corroborated by the increasing $\rho$ from $lbt_{30\%M}$ to $lbt_{90\%M}$. The color-coding suggests that it is more difficult to infer SFH of galaxies with high sSFR than the galaxies with low sSFR. This may be in part caused by the outshining effect of young bright stars, which is abundant in the galaxies with high sSFR.  

By fitting synthetic FUV--FIR photometric SEDs of simulated galaxies using a modified version of \texttt{MAGPHYS}, which assumes tau SFH model with random bursts superposed, \citet{2015Smith} tried to reconstruct the true SFH of the galaxies. They found that the median-likelihood SFH (obtained by marginalizing over the model libraries, which is similar to what is done in our work) can well recover the smoothly declining SFH of isolated disk galaxies, while it fails to recover the bursty episodes in the SFH of merging galaxies. This is likely caused by the assumed tau SFH model that is not flexible enough to represent the general (i.e.,~realistic) SFH of galaxies. Inline with our results, \citet{2018Carnall} showed that using the more flexible double power law SFH model can recover the overall shape of the true SFHs of simulated galaxies from \texttt{MUFASA}, despite the narrower wavelength coverage (optical--NIR) of the mock SEDs used. While \verb|piXedfit_fitting| module can recover the overall trend of rising and falling episodes (i.e.,~low frequency variation) in the true SFHs of TNG galaxies, however, it cannot recover the high frequency variation in the true SFHs.   

\begin{figure*}[ht]
\centering
\includegraphics[width=0.9\textwidth]{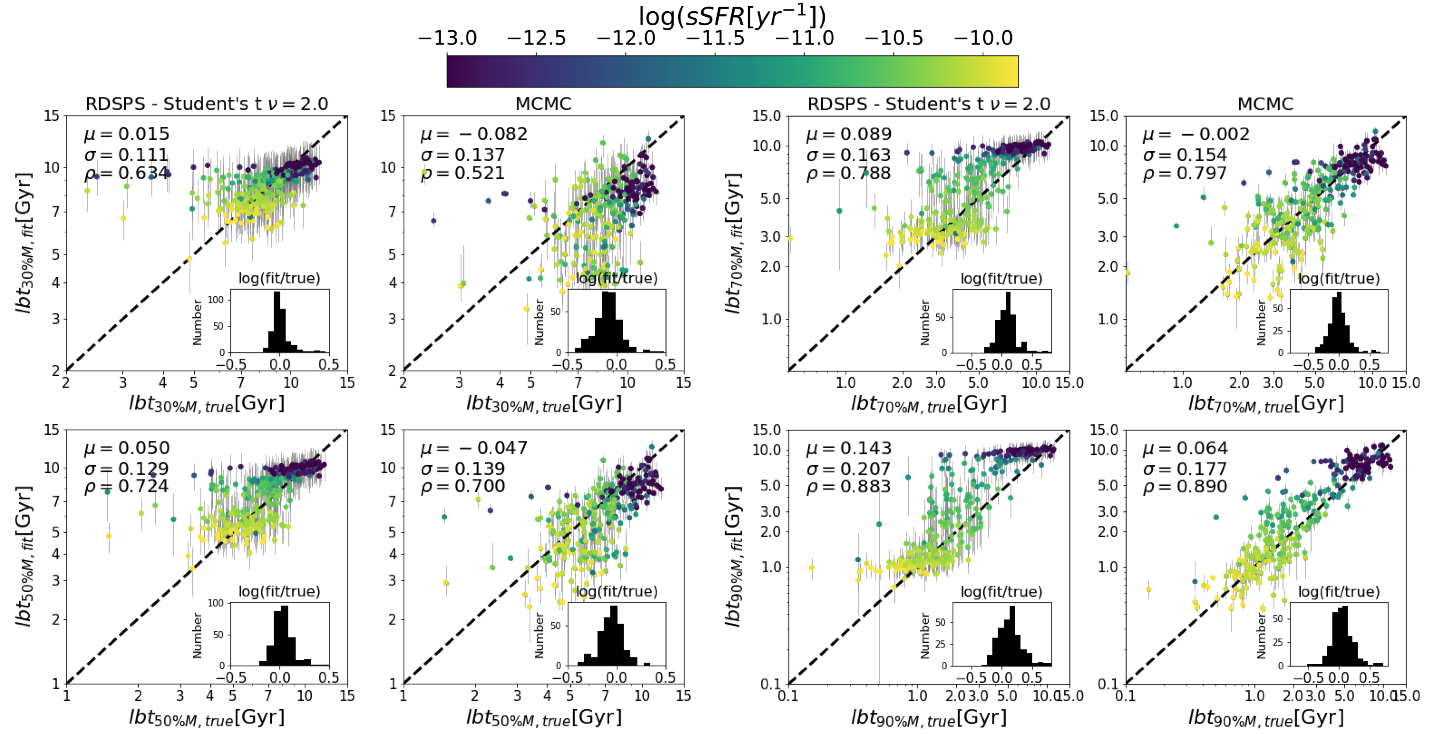}
\caption{Similar to Figure~\ref{fig:recov_SFH_lbt_photo}, but now the fitting is done to the spectrophotometric SEDs of the TNG galaxies.}
\label{fig:recov_SFH_lbt_specphoto}
\end{figure*}

It is interesting to see how the inclusion of the spectrum into the SED can effect the SFH inference. Figure~\ref{fig:recov_SFH_lbt_specphoto} shows comparison between the inferred $lbt_{30\%M}$, $lbt_{50\%M}$, $lbt_{70\%M}$, and $lbt_{90\%M}$ obtained from fitting to the mock spectrophotometric SEDs. The format of this figure is the same as that of Figure~\ref{fig:recov_SFH_lbt_photo}. Overall, we see improvements made by the fitting with the spectrophotometric SED, such that the scatters in the one-to-one comparisons become smaller and the $\rho$ values become higher compared to that obtained from the fitting with photometric SED only. However, the offsets become slightly higher. We notice a flattening appears around the highest and lowest ends of the correlation in case of the fitting with the RDSPS approach. This flattening can be caused by a multimodal posteriors distributions.

\section{Testing the Performance of \texttt{piXedfit} Using Spatially Resolved Spectrophotometric Data of Local Galaxies} \label{sec:test_with_specphoto_data}
In this section, we analyze spatially resolved spectrophotometric data of 10 galaxies observed by the CALIFA survey and 10 galaxies observed by the MaNGA survey. The goals of this analysis are (1) to demonstrate the ability of the \verb|piXedfit| in spatially matching the FUV--$W2$ broad-band imaging data and the IFS data, (2) demonstrate the SED fitting analysis to the spatially resolved spectrophotometric dataset with \verb|piXedfit|, and (3) test the reliability of the SED fitting module by comparing the inferred SFR from fitting with the SFR derived from $H_{\alpha}$ emission. A more scientific-oriented discussion using larger samples is left for future works.    

\subsection{Sample Selection and Data Reduction} \label{sec:specphoto_sample_select}
First, we construct a catalog of galaxies that are observed by the medium imaging survey (MIS) of GALEX, SDSS, 2MASS, and WISE. We start from the catalog of unique GALEX GR5 sources (i.e.,~eliminating repeated measurements) that has been matched with the SDSS DR7 catalog by \citet{2011Bianchi}\footnote{Available at \url{http://dolomiti.pha.jhu.edu/uvsky}}, then cross match it with the MPA-JHU (Max-Planck-Institut f\"ur Astrophysik-Johns Hopkins University) value added galaxy catalog\footnote{Available at \url{https://wwwmpa.mpa-garching.mpg.de/SDSS/DR7/}} \citep{2003Kauffmann, 2004Tremonti, 2004Brinchmann} to select only galaxies and get their stellar masses. After that we cross match the catalog with the 2MASS extended source catalog\footnote{Available at \url{https://irsa.ipac.caltech.edu/Missions/2mass.html}}\citep{2000Jarrett}. Considering the all sky coverage of the WISE survey and its depth compared to 2MASS, we do not cross match the catalog further with the WISE catalog. Once we get the catalog, we cross match it with the CALIFA DR3\footnote{Available at \url{https://califaserv.caha.es/CALIFA_WEB/public_html/?q=content/califa-3rd-data-release}} and MaNGA DRPALL (from SDSS DR15\footnote{Available at \url{https://www.sdss.org/dr15/manga/manga-data/catalogs/}}) catalogs separately. As a result we get 41 galaxies matched with the CALIFA catalog and 395 galaxies matched with the MaNGA catalog. Then we randomly select 10 galaxies that have $\log(M_{*}[M_{\odot}])>10.0$ and $z<0.05$ from each of the two catalogs. We download the multiband images and the IFS data from the relevant survey websites, assisted by the galaxies coordinates from the merged catalog.  

We require the galaxies to be covered by the GALEX MIS because the survey has relatively long exposure time (typically $1500$ s) so that we can have sufficient S/N ratio in the UV. Among the imaging datasets used, the 2MASS imaging data is the shallowest. However, it is still important to include the data because it complements the two WISE bands in putting strong constraint in the NIR regime. In total, the photometry data consists of 12 bands ranging from FUV to $W2$. We use the \verb|piXedfit_images| module to spatially match (in resolution and sampling) the imaging data, then use \verb|piXedfit_spectrophotometric| module to spatially match the reduced imaging data cubes with the IFS data. This processes produce  spatially resolved spectrophotometric data cubes that have spatial resolution similar to that of the $W2$ and spatial sampling similar to that of the FUV/NUV.  

\begin{figure*}[ht]
\centering
\includegraphics[width=0.8\textwidth]{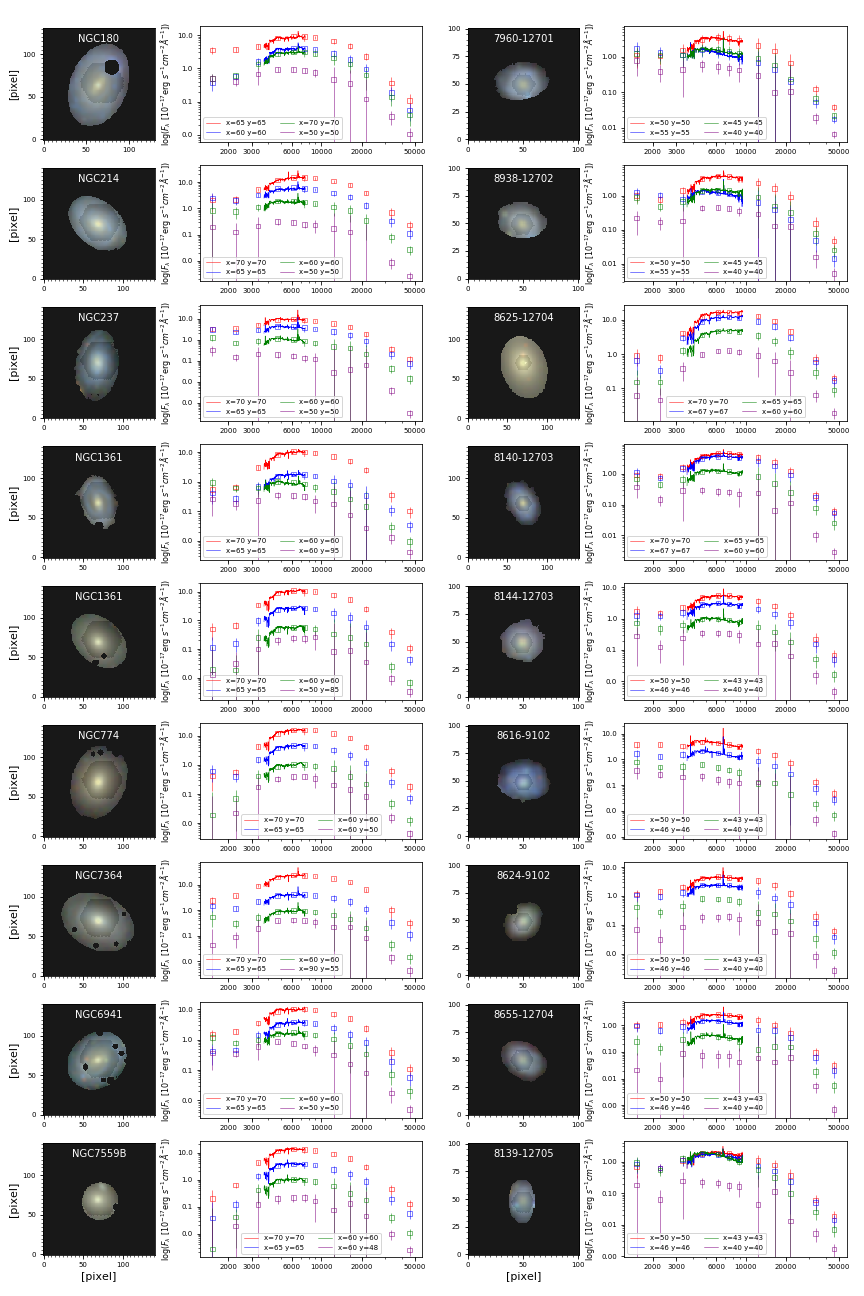}
\caption{Compilation of the spatially resolved spectrophotometric data cubes of 18 galaxies from the sample analyzed in this paper. The data cubes of the other 2 galaxies are shown in Figure~\ref{fig:specphotoSED_califa_manga1}. The left and right sides show 9 galaxies from the CALIFA and 9 galaxies from the MaNGA, respectively. For each galaxy, $gri$ composite image (left panel) and SEDs of four pixels (right panel) are shown. In the $gri$ composite images, the area covered by the IFU fiber bundle is shown with the transparent hexagonal area.}
\label{fig:plot_collect_specphoto}
\end{figure*}

Figure~\ref{fig:plot_collect_specphoto} shows spatially resolved spectrophotometric data cubes of 18 galaxies from the sample. The data cubes of the other 2 galaxies are shown in Figure~\ref{fig:specphotoSED_califa_manga1}. In the left side, the 9 galaxies from CALIFA are shown, while the 9 galaxies from MaNGA are shown in the right side. For each galaxy, $gri$ composite image and example of SEDs of four pixels are shown in the left panel and the right panel, respectively. In the $gri$ composite image, the transparent hexagonal area shows the area covered by the IFU fiber bundle of the CALIFA and MaNGA surveys. In the data cubes, only pixels covered within the region of the IFU fiber bundle have the spectra, while pixels outside of the region only have photometric SED.  

The spectrophotometric data cubes are then passed to the pixel binning process. The pixel binning is done using the \verb|piXedfit_bin| module. See Section~\ref{sec:piXedfit_bin} for the description of the pixel binning scheme. In this analysis, the criteria for the binning are: S/N ratio threshold of $10.0$ in the GALEX, SDSS, and WISE bands, S/N ratio threshold of $1.0$ in the 2MASS bands, $D_{\rm min,bin}$ of $4$ pixels, and reduced $\chi^{2}$ limit ($\chi^{2}_{\rm max,bin}$) of $3.3$ for the SED shape similarity test. In short, the pixel binning is done by growing the size of bins and including more pixels with similar SED shape until the S/N thresholds in all bands are achieved. Pixel binning maps of the 20 galaxies analyzed in this work are shown in the leftmost panels of Figures~\ref{fig:plot_collect_maps_props_califa} and~\ref{fig:plot_collect_maps_props_manga}.

\subsection{SED Fitting Analysis} \label{sec:example_fitting}
The reduced spectrophotometric data cubes (after pixel binning) are then passed to the SED fitting process. The SED fitting is done using the \verb|piXedfit_fitting| module (see Section~\ref{sec:SEDfit_procedure}) with MCMC approach. The SED fitting setup (IMF, isochrone, spectral library, SFH, and dust attenuation law) is the same as that for the fitting with the mock SEDs of the TNG galaxies (see Section~\ref{sec:SEDfit_analysis_TNG}), except for the priors. Flat priors for all parameters are assumed, within the following ranges: $\log(Z/Z_{\odot})=[-2.0,0.2]$, $\log(\tau)=[-1.0,1.14]$, $\log(\text{age})=[-1.0,1.14]$, $\hat{\tau}_{2}=[0.0,3.0]$, $\log(\alpha)=[-2.0,1.5]$, and $\log(\beta)=[-2.0,1.5]$. The $M_{*}$ prior is defined in the same way as that applied in the fitting with the mock SEDs of TNG galaxies. The reason of using different set of priors from those used for fitting the mock SEDs of the TNG galaxies is because here we analyze the spatially resolved SEDs, which come from stellar populations with wide range of ages ($\text{age}_{\text{sys}}$). In the MCMC fitting, we set the number of walkers and steps as 100 and 1000, respectively.

For spatial bins that have spectrophotometric SED (see Section~\ref{sec:piXedfit_bin} for the definition of spatial bins with spectrophotometric SEDs), two kinds of fitting are done: a fitting to the photometric SED only and a fitting to the spectrophotometric SED. By default, \verb|piXedfit_fitting| will fit both photometric SED and spectrum (i.e.,~the spectral continuum) simultaneously whenever it is fed with a spectrophotometric data. The aim of performing fitting to only the photometric SED is for conducting tests, which includes a reconstruction of the observed spectral continuum, $\text{D}_{\rm n}4000$, $H_{\alpha}$ emission, and $H_{\beta}$ emission using model spectra obtained from fitting to the photometry. These analyses will be discussed in the next two sections.
    
\begin{figure*}[ht]
\centering
\includegraphics[width=1.0\textwidth]{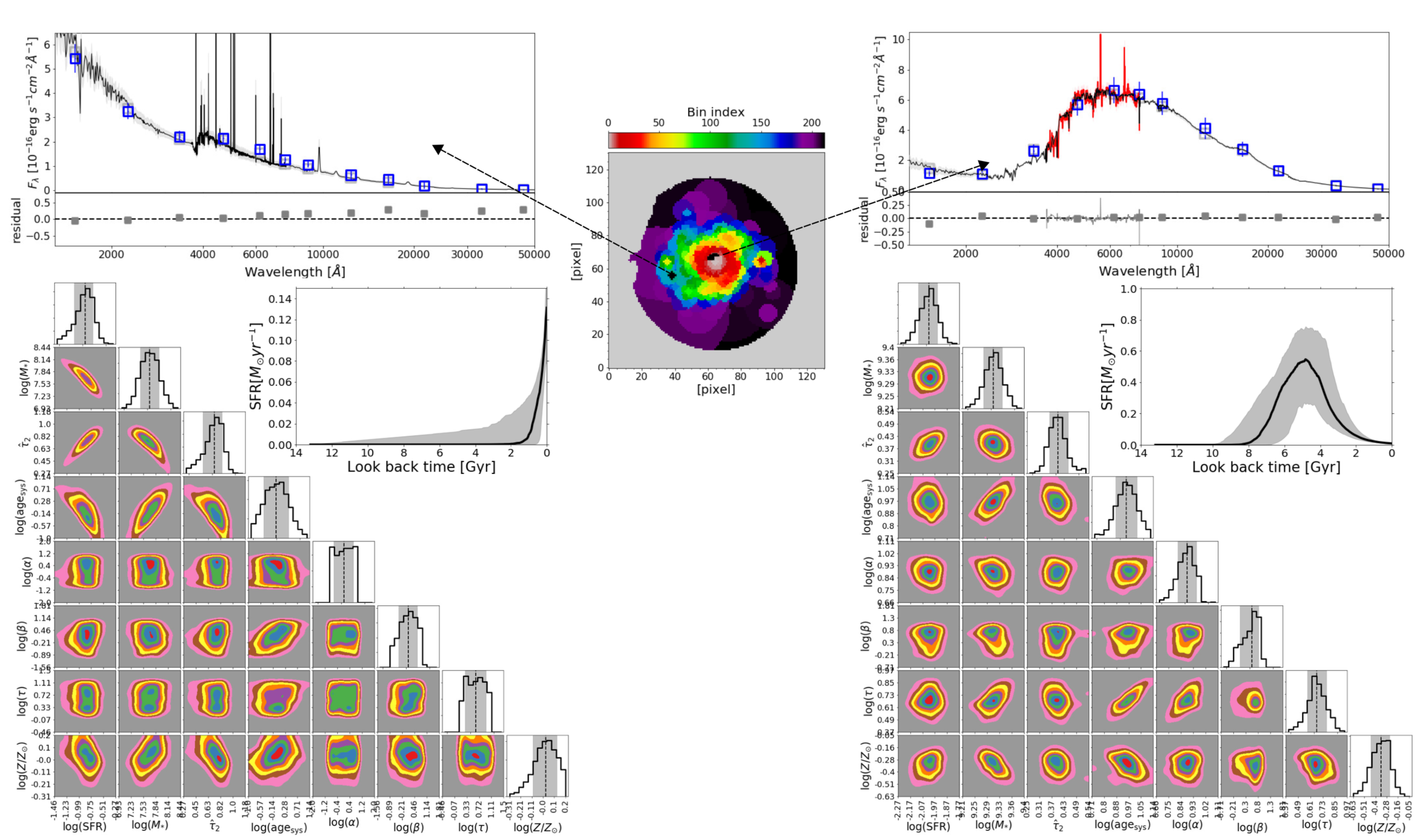}
\caption{Examples of fitting results using MCMC for two spatial bins of the NGC 309, one located around the center and the other located in the spiral arms. The fitting to the centrally-located bin is done to the spectrophotometric SED, while the fitting for the outter bin is done to the photometric SED. The overall symbols in the corner plot, SED plot, and SFH plot are the same as that in the Figure~\ref{fig:demo_pixedfit_tng}.}
\label{fig:demo_pixedfit_ngc309}
\end{figure*} 

Figure~\ref{fig:demo_pixedfit_ngc309} shows fitting results using MCMC for two spatial bins in the NGC 309, one located around the center and the other located in the spiral arms. For centrally-located bin, the fitting is done to both spectrum and photometric SED simultaneously. The overall symbols in the corner plot, SED plot, and SFH plot are the same as that in the Figure~\ref{fig:demo_pixedfit_tng}. An obvious difference in the SED shape between the spatial bin around the galaxy's center (a red SED typical of old stellar population) and that in the spiral arms (a blue SED typical of young stellar population) is shown in the SED plots. In the SED plot, the black spectrum and gray shaded area around it represent median posterior model spectrum and the associated uncertainty. The small residuals in the SED plot indicate that the observed continuum and photometric SED can be recovered well. The inferred SFH of the spatial bin located in the spiral arms indicates a steeply increasing SFR toward the observational time, while the inferred SFH of the spatial bin located around the galaxy's center indicates a gradual increase of SFR from $\sim10$ Gyr ago and reached peak around $\sim5$ Gyr ago then the SFR gradually decrease toward the observational time.     

The SED fitting procedure is done to the SEDs of spatial bins in the galaxy, then for parameters that linearly scaled with flux at a certain band, such as $M_{*}$ (which scaled with NIR bands) and SFR (which scaled with UV bands), the inferred value of the bin is divided into the pixels that belong to the bin by assuming that $M_{*}$ is proportional to the $W2$ flux and SFR is proportional to the FUV flux. This way, we can get higher spatial resolution in the $M_{*}$ and SFR maps. The other parameters are kept in the spatial bin space. 

\begin{figure*}
\centering
\includegraphics[width=0.9\textwidth]{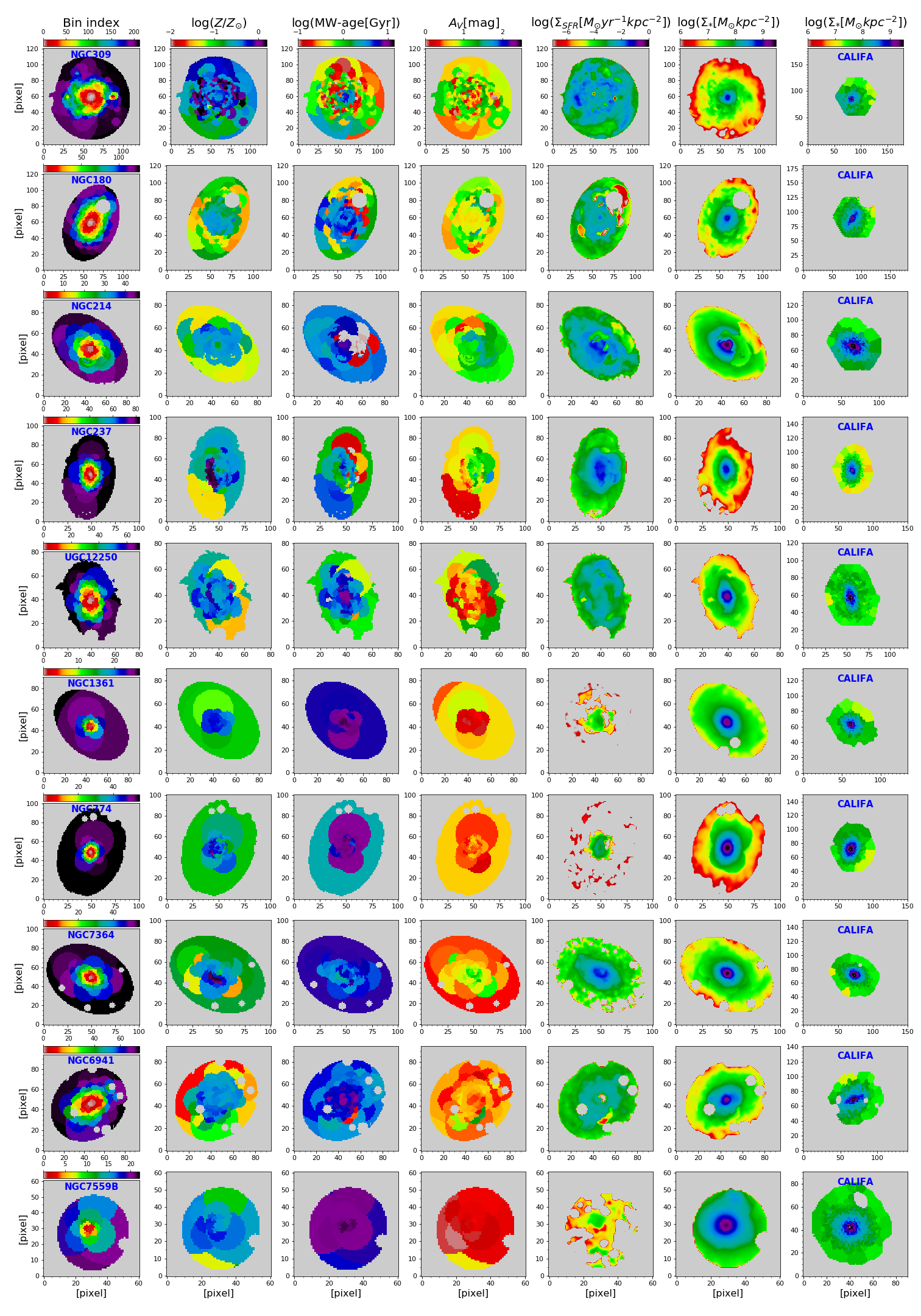}
\caption{Maps of the stellar population properties of the 10 galaxies from the CALIFA survey analyzed in this work. The SED fitting uses the MCMC technique. The first 6 columns from the left show maps of the pixel binning, $Z$, mass-weighted age, $A_{V}$ dust attenuation, $\Sigma_{\rm SFR}$, and $\Sigma_{*}$. The righmost column show $\Sigma_{*}$ from the PyCASSO data base \citep{2017deAmorim} which is derived from the CALIFA data alone. The dimension of this map corresponds to the same physical size as those of the dimensions of the maps in the other 6 columns.}
\label{fig:plot_collect_maps_props_califa}
\end{figure*}

\begin{figure*}
\centering
\includegraphics[width=0.9\textwidth]{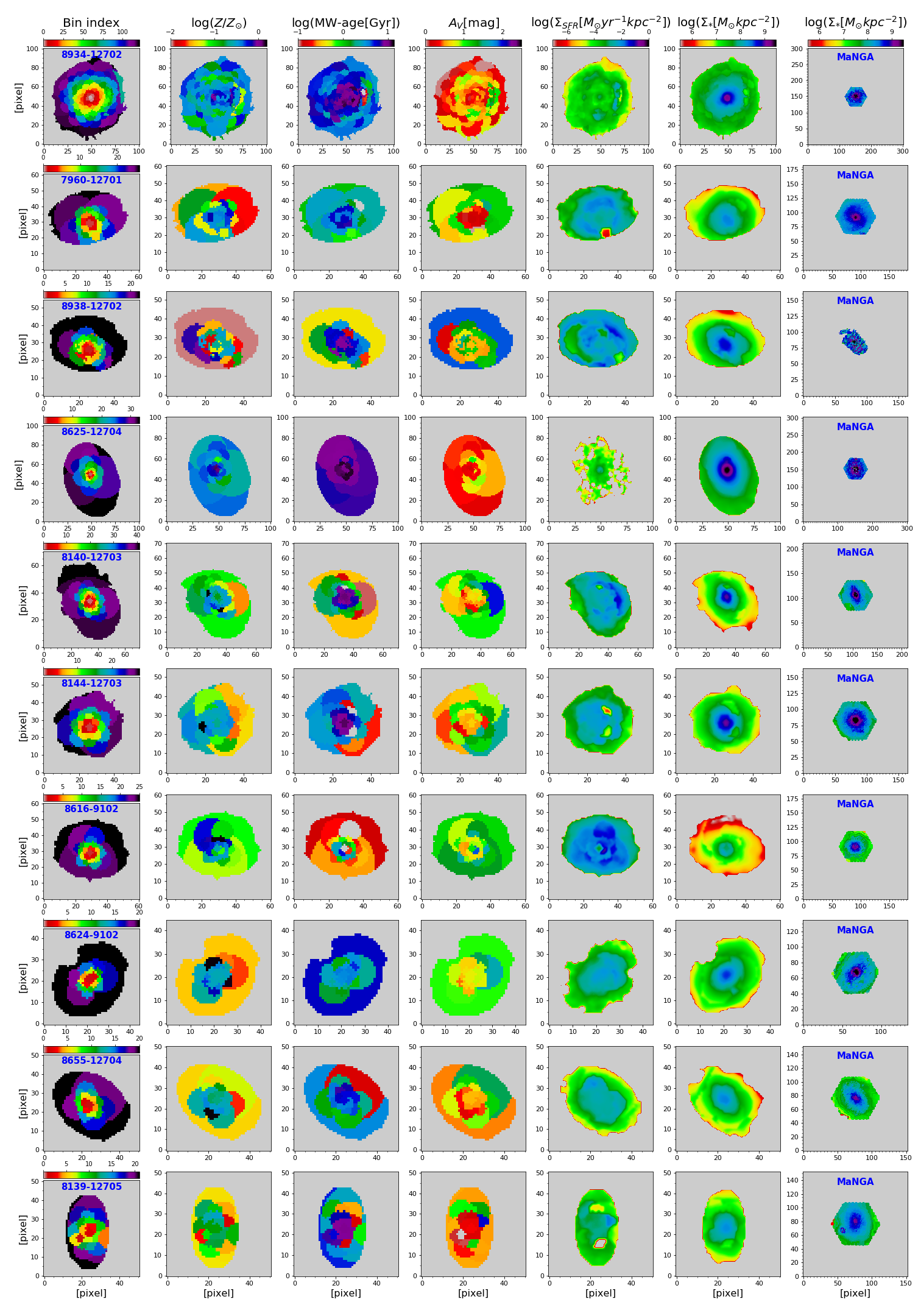}
\caption{Maps of the stellar population properties of the 10 galaxies from the MaNGA survey analyzed in this work. The SED fitting uses MCMC technique. The first 6 columns from the left show maps of the pixel binning, $Z$, mass-weighted age, $A_{V}$ dust attenuation, $\Sigma_{\rm SFR}$, and $\Sigma_{*}$. The righmost column show $\Sigma_{*}$ from the Pipe3D value added catalog \citep{2018Sanchez} which is based on MaNGA data only. The dimension of this map corresponds to the same physical size as those of the dimensions of the maps in the other 6 columns.}
\label{fig:plot_collect_maps_props_manga}
\end{figure*}

Figures~\ref{fig:plot_collect_maps_props_califa} and~\ref{fig:plot_collect_maps_props_manga} show the maps of stellar population properties of the 10 galaxies from CALIFA and the 10 galaxies from MaNGA, respectively. In the both figures, the first 6 columns from the left show maps of the pixel binning, $Z$, mass-weighted age, $A_{V}$ dust attenuation, SFR surface density ($\Sigma_{\rm SFR}$), and $M_{*}$ surface density ($\Sigma_{*}$). The mass-weighted age is derived from the inferred SFH, while the $A_{V}$ can be calculated as $A_{V}=1.086\times \hat{\tau}_{2}$. For comparison, in the rightmost columns of the both figures, we show $\Sigma_{*}$ from the PyCASSO data base\footnote{ Available at \url{http://pycasso.iaa.es/}} \citep{2017deAmorim} which is derived from the CALIFA data alone (in case of figure~\ref{fig:plot_collect_maps_props_califa}) and $\Sigma_{*}$ from the Pipe3D value added catalog\footnote{Available at \url{https://www.sdss.org/dr14/manga/manga-data/manga-pipe3d-value-added-catalog/}} \citep{2018Sanchez} which is based on the MaNGA data alone (in case of figure~\ref{fig:plot_collect_maps_props_manga}). The dimensions of the plotted maps in these rightmost columns correspond to the same physical sizes as those of the dimensions of the maps in the other 6 columns. The fact that our data cubes have lower spatial sampling (i.e.,~larger pixel size; $1.5''\text{ pixel}^{-1}$) than that of the CALIFA ($1.0''\text{ pixel}^{-1}$) and MaNGA ($0.5''\text{ pixel}^{-1}$) data cubes makes our maps have smaller total number of pixels.   

\subsection{Reconstructing Observed Spectral Continuum with Model Spectra Obtained from Fitting to Photometry} \label{sec:spec_cont_reconstruct}
In this section and the next section, for spatial bins that have spectrophotometric SEDs, we fit the photometric SEDs and then compare the median posterior model spectra with the observed spectra (see Section~\ref{sec:SEDfit_analysis_TNG} for the description on how the median posterior model spectra are obtained). We make the comparison by calculating the residual in spectral continuum (in this section) and directly comparing the $\text{D}_{\rm n}4000$ strength, and the $H_{\alpha}$ and $H_{\beta}$ luminosities (in the next section). This analysis can serves as an excellent test for \verb|piXedfit| in terms of its SED modeling (based on FSPS) and the fitting performance. A similar exercise has been carried out by \citet{2017Leja} with the \verb|prospector|, but for galaxies as a whole.

\begin{figure*}[ht]
\centering
\includegraphics[width=0.7\textwidth]{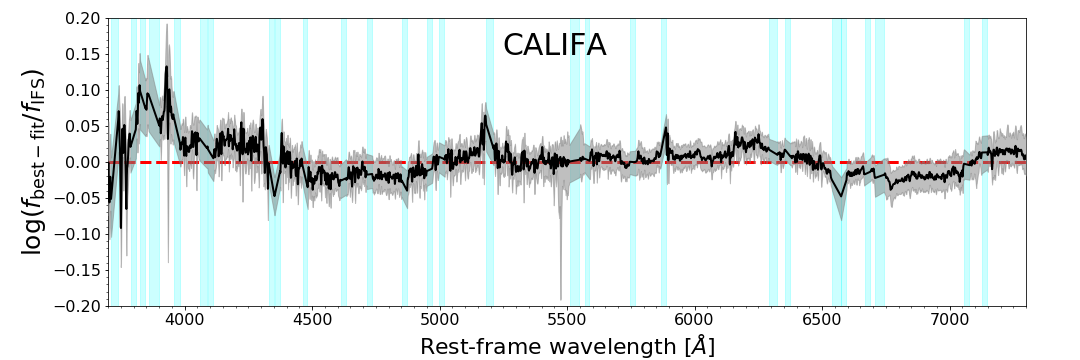}
\includegraphics[width=0.7\textwidth]{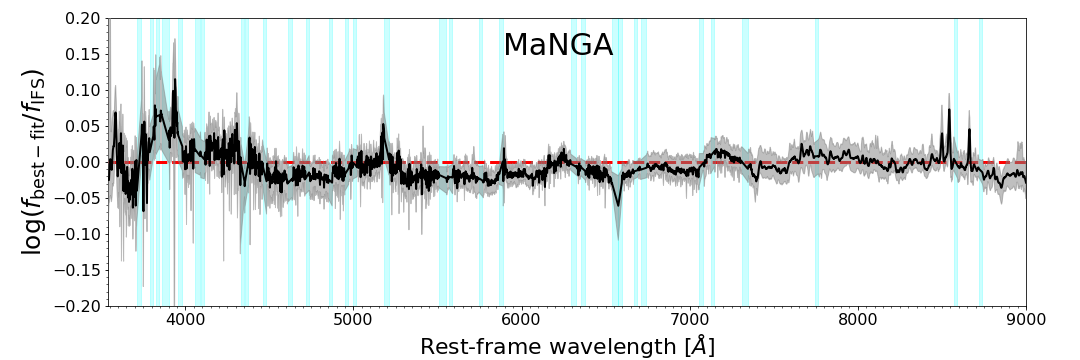}
\caption{Merged residuals between the median posterior model spectra obtained from fitting to the photomeric SEDs and the observed spectra for the CALIFA (top panel) and MaNGA (bottom panel) sample. The black lines and gray shaded areas are the median and 16th--84th percentiles of the residuals. The vertical cyan bands in the two panels show regions in the spectra that are removed.}
    \label{fig:califa_manga_merge_flux_ratio_bfit_ifu}
\end{figure*}

We collected the spatial bins that have spectrophotometric SEDs in the CALIFA (560 bins) and MaNGA (145 bins) samples. To get the continuum from the observed spectra and the median posterior spectra, we remove regions within $\pm 10\text{\normalfont\AA}$ from the central wavelengths of all possible emission lines (based on the list of emission lines wavelengths from the FSPS). Figure~\ref{fig:califa_manga_merge_flux_ratio_bfit_ifu} shows the residuals (in dex) between the observed spectra and median posterior model spectra obtained from fitting to the photometric SEDs of pixels in the CALIFA (top panel) and MaNGA (bottom panel) samples. The merged residuals are brought to the rest-frame wavelength. The black lines show the median of the residuals, while the gray shaded areas show 16th--84th percentiles. The vertical cyan bands in the two panels show regions in the spectra that are removed. The residuals are relatively flat over the whole wavelength ranges. The mean and standard deviation of the residuals are $0.004$ and $0.037$ (for CALIFA) and $-0.005$ and $0.031$ (for MaNGA). Larger residuals are shown around the $4000\text{\normalfont\AA}$ break. This could be caused by the lack of photometric sampling around that region, which currently only covered by the $u$ and $g$ bands.

\subsection{Predicting $H_{\alpha}$, $H_{\beta}$, and $\text{D}_{\rm n}4000$ with Model Spectra Obtained from Fitting to Photometry} \label{sec:spec_features_reconstruct}
In this section, we try to predict $H_{\alpha}$ luminosity, $H_{\beta}$ luminosity, and $\text{D}_{\rm n}4000$ of the observed spectra through fitting using \verb|piXedfit_fitting| with only photometric SED. 
%For this analysis, we collected all the spatial bins from the CALIFA (560 bins) and MaNGA (145 bins) samples that have spectrophotometric SED. 
To measure luminosities of the $H_{\alpha}$ and $H_{\beta}$ emission lines from the observed spectra, first, we subtract the observed spectra with the continuum of the median posterior model spectra, generated from the model posteriors. Then we fit the $H_{\alpha}$ and $H_{\beta}$ emission lines with Gaussian functions using \verb|fit_lines| function in the \verb|specutils| \citep{2020nicholas}\footnote{\url{https://specutils.readthedocs.io/en/stable/}} Python package.  
We visually inspect all the spatial bins to make sure the fitting work well. 
Uncertainties of the $H_{\alpha}$ and $H_{\beta}$ luminosities are estimated based on the average $\text{S}/\text{N}$ ratio of the observed spectral fluxes $\pm 10\text{\normalfont\AA}$ around the mean wavelengths.  
The $H_{\alpha}$ and $H_{\beta}$ luminosities of the median posterior model spectra are derived from the posteriors distributions obtained from the MCMC fitting. The median, 16th, and 84th percentiles are calculated from the posteriors distributions. The median is then used as the mean luminosity, while the 16th--84th percentiles are used as the uncertainty.

The $\text{D}_{\rm n}4000$ of both the observed spectra and the median posterior model spectra is measured following the \citet{1999Balogh} definition, which is the ratio of the average flux density $f_{\nu}$ in the narrow wavelength bands of $3850$--$3950\text{\normalfont\AA}$ and $4000$--$4100\text{\normalfont\AA}$. To estimate uncertainty for the $\text{D}_{\rm n}4000$ of the observed spectra, we use the bootstrap method; The spectral fluxes within the two bands are randomly perturbed following a Gaussian distribution with the mean of the spectral fluxes and the standard deviation of the flux uncertainties. This is performed 100 times and for each iteration, $\text{D}_{\rm n}4000$ is measured. Then from the distribution, a standard deviation is calculated and used as the $\text{D}_{\rm n}4000$ uncertainty.  

\begin{figure*}
\centering
\includegraphics[width=0.9\textwidth]{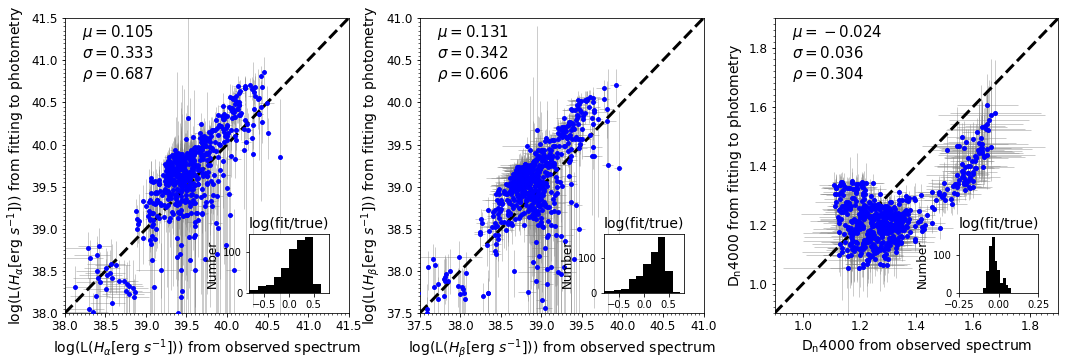}
\includegraphics[width=0.9\textwidth]{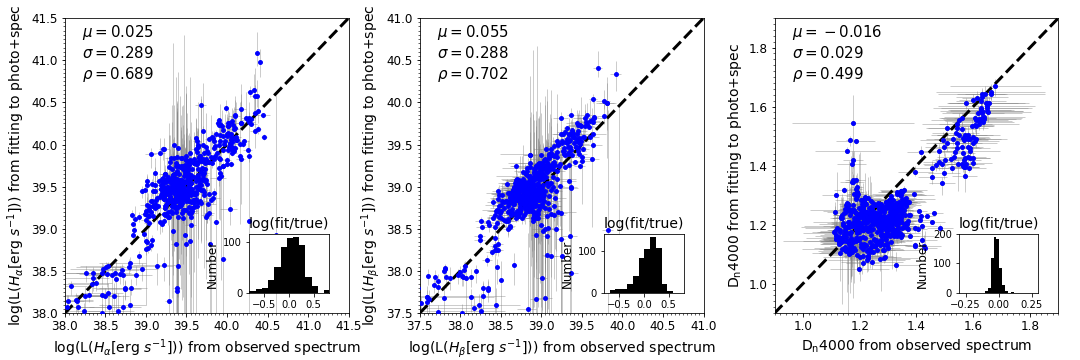}
\caption{Comparison between the observed $H_{\alpha}$ luminosities, $H_{\beta}$ luminosities, and $\text{D}_{\rm n}4000$, and the model predictions from fitting to photometric SEDs (first row) and spectrophotometric SEDs (second row). In each panel, the histogram in the bottom right corner shows distribution of the logarithmic ratio between the models and the observed ones. The offset ($\mu$), scatter ($\sigma$), and Spearman rank-order correlation coefficient ($\rho$) are shown in the top left corner.}
\label{fig:comp_HaHbD4000_obs_bfit_mcmc}
\end{figure*}

In Figure~\ref{fig:comp_HaHbD4000_obs_bfit_mcmc}, first row, we show comparison between the observed $H_{\alpha}$ luminosity (left), $H_{\beta}$ luminosity (middle), and $\text{D}_{\rm n}4000$ (right) and the predictions by \verb|piXedfit_fitting| through fitting with photometric SEDs. In each panel, the histogram in the bottom right corner shows distribution of the logarithmic ratio between the model predictions and the observed ones. The mean (or offset in dex, $\mu$), scatter ($\sigma$), and Spearman rank-order coefficient ($\rho$) of the distribution are shown in the top left corner. There is a good agreement between the models and observations, especially for the $H_{\alpha}$ and $H_{\beta}$, corroborated by the small offsets ($0.105$ and $0.131$ dex for the $H_{\alpha}$ and $H_{\beta}$, respectively) and scatters ($0.333$ and $0.342$ dex for the $H_{\alpha}$ and $H_{\beta}$, respectively). The $\rho$ values for $H_{\alpha}$ and $H_{\beta}$ are relatively high ($0.678$ and $0.606$, respectively), confirming the good agreement between the models and observations. Despite the small offset ($-0.024$ dex) and scatter ($0.036$ dex), however, the $\rho$ values for $\text{D}_{\rm n}4000$ is small ($0.304$) which is caused by the deviation (from the one-to-one relation) around the intermediate $\text{D}_{\rm n}4000$ and larger scatter around the lower end of the $\text{D}_{\rm n}4000$. This trend in the $\text{D}_{\rm n}4000$ comparison is understandable given the non-flat residuals around the $4000\text{\normalfont\AA}$ between the predicted spectral continuum and the observed spectral continuum as shown in Figure~\ref{fig:califa_manga_merge_flux_ratio_bfit_ifu}.

In order to see whether including the observed spectral continuum in the fitting can improve the model predictions, we fit the spectrophotometric SEDs of the spatial bins and derived the model predictions for the $H_{\alpha}$, $H_{\beta}$, and $\text{D}_{\rm n}4000$ with the same procedure as described previously. The comparisons between the model predictions and the observed ones are shown in the second row of Figure~\ref{fig:comp_HaHbD4000_obs_bfit_mcmc}. A better agreement between the models and the observations is obtained with this fitting compared to the previous one that only use photometric SEDs. This is indicated by smaller offsets, smaller scatters, and higher $\rho$ values in all of the three comparisons.                    

\subsection{Comparison of SFR from \texttt{piXedfit\_fitting} with the SFR Derived from $H_{\alpha}$} \label{sec:comp_bfit_SFR_SFR_Halpha} 
The Balmer emission lines, especially $H_{\alpha}$ line (which is the strongest) is a good indicator of instantaneous SFR. In addition to that, the Balmer decrement (i.e.,~ratio of $H_{\alpha}/H_{\beta}$ emission line fluxes) provides a good indicator for dust attenuation in the stars birth clouds. The $H_{\alpha}$-based SFR estimate has been widely used in the analysis of the IFS data in the CALIFA \citep[e.g.,][]{2016Sanchez_b} and MaNGA \citep[e.g.,][]{2018Sanchez, 2019Belfiore} surveys. 

In this section, for spatial bins with spectrophotometric SEDs and the $H_{\alpha}$ and $H_{\beta}$ $\text{S}/\text{N}>2.0$ (resulting in 527 bins), we compare the SFR derived with \verb|piXedfit_fitting| module and the SFR derived from the observed $H_{\alpha}$ emission. Here, we use the $H_{\alpha}$ and $H_{\beta}$ measurements from the analysis in the previous section (Section~\ref{sec:spec_features_reconstruct}). We do not use the publicly available value added data cubes from the CALIFA and MaNGA surveys, because of the differences in spatial resolution and spatial sampling between their data cubes and our reduced data cubes. Spatially matching the SFR map or $H_{\alpha}$ and $H_{\beta}$ maps from their data cubes to our data cubes will introduce some systematics that could dominate uncertainties in the comparison analysis. Moreover, the comparison can be made more self-consistent.

In deriving SFR from the $H_{\alpha}$ emission, first we correct the $H_{\alpha}$ luminosity for the dust attenuation associated with the birth cloud. Balmer color excess is correlated with the ratio of the observed Balmer decrement $(L_{H_{\alpha}}/L_{H_{\beta}})_{\rm obs}$ and its intrinsic value $(L_{H_{\alpha}}/L_{H_{\beta}})_{\rm int}$ through the following equation:
\begin{equation} 
E(H_{\beta}-H_{\alpha}) = 2.5\log \left(\frac{(L_{H_{\alpha}}/L_{H_{\beta}})_{\rm obs}}{(L_{H_{\alpha}}/L_{H_{\beta}})_{\rm int}} \right).
\end{equation} 
The $L_{H_{\alpha}}$ and $L_{H_{\beta}}$ are the luminosities of $H_{\alpha}$ and $H_{\beta}$, respectively. The intrinsic Balmer decrement has a value of $2.86$ for the case B recombination \citep{1989Osterbrock}. Once we have the Balmer color excess, the attenuation toward $H_{\alpha}$ can then be calculated as: 
\begin{equation}
A_{H_{\alpha}} = \frac{E(H_{\beta} - H_{\alpha})}{k(\lambda_{H_{\beta}}) - k(\lambda_{H_{\alpha}})} \times k(\lambda_{H_{\alpha}}). 
\end{equation}
The $k(\lambda_{H_{\alpha}})$ and $k(\lambda_{H_{\beta}})$ are the attenuation values at wavelengths of $H_{\alpha}$ and $H_{\beta}$, respectively. To get these values, we assume the \citet{2000Calzetti} attenuation curve with $R_{V}=3.1$. It is important to note that \citet{2000Calzetti} used two different attenuation curves for the nebular and continuum. The two attenuation curves have similar shapes but different normalizations: $R_{V}=3.1$ and $R_{V}=4.05$ for the nebular and continuum, respectively. Once we have $A_{H_{\alpha}}$, the dust-corrected $H_{\alpha}$ luminosity can then be calculated via:
\begin{equation}
L_{H_{\alpha},\text{corr}} = L_{H_{\alpha},\text{obs}} \times 10^{0.4A_{H_{\alpha}}} 
\end{equation}   

For deriving the SFR from the dust-corrected $H_{\alpha}$ luminosity, we use the \citet{1998Kennicutt_b} prescription that has been converted for \citet{2003Chabrier} IMF as follows:
\begin{equation}
\text{SFR}[M_{\odot}\text{yr}^{-1}] = 4.65 \times 10^{-42}L_{H_{\alpha},\text{corr}}[\text{erg}\text{ s}^{-1}].
\end{equation} 
A division by $1.7$ has been applied to the original \citet{1998Kennicutt_b} prescription (which assumed \citealt{1955Salpeter} IMF) to account for additional low-mass stars in the Salpeter IMF \citep[see e.g.,][]{2014Speagle, 2016Nelson, 2017Leja}. The uncertainty of SFR from $H_{\alpha}$ is estimated using the bootstrap method. 

\begin{figure}
\centering 
\includegraphics[width=0.35\textwidth]{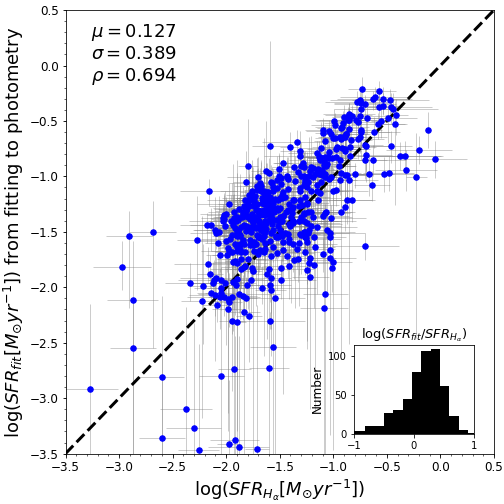}
\includegraphics[width=0.35\textwidth]{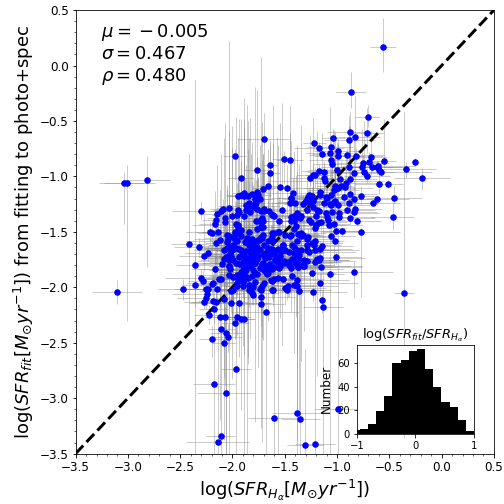}
\caption{Comparison between the SFR derived from fitting with the MCMC method and the SFR derived from $H_{\alpha}$ emission. In the top panel, the fitting is done to the photometric SEDs of spatial bins, while in the bottom panel, the fitting is done to the spectrophotometric SEDs. In each panel, the histogram in the bottom right corner shows distribution of the logarithmic ratio between the two SFR estimates. The offset ($\mu$), scatter ($\sigma$), and Spearman $\rho$ values are shown in the top left corner.}
\label{fig:SFRpixedfit_vs_SFRHalpha_mcmc}
\end{figure}

Figure~\ref{fig:SFRpixedfit_vs_SFRHalpha_mcmc}, top panel, shows comparison between the SFR obtained from fitting to the photometric SEDs of spatial bins and the SFR derived from the $H_{\alpha}$ emission. There is a good agreement between the two SFR measurements, with a small offset of $0.127$ dex and scatter of $0.389$ dex. The Spearman $\rho$ value is high ($0.694$), confirming the good agreement between the two SFR measurements. To see whether including spectral continuum in the fitting can make an improvement to the result, we fit the spectrophotometric SEDs of the spatial bins. The result is shown in the bottom panel of Figure~\ref{fig:SFRpixedfit_vs_SFRHalpha_mcmc}. Now, the offset is significantly reduced to $-0.005$ dex. However, the scatter becomes slightly larger ($0.467$ dex) and the Spearman $\rho$ value is reduced to $0.480$.        

\begin{figure}
\centering
\includegraphics[width=0.35\textwidth]{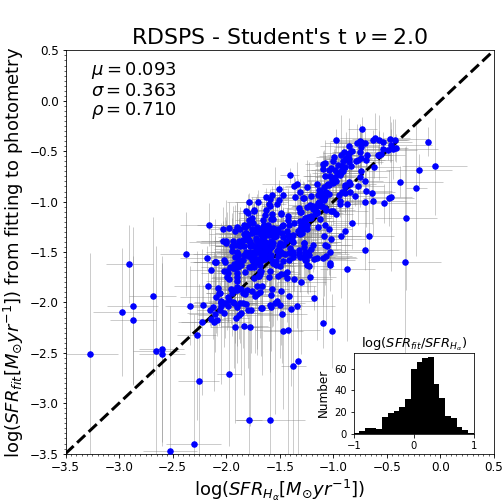}
\caption{Comparison between the SFR derived from $H_{\alpha}$ emission and SFR from fitting using the RDSPS method with the likelihood function of Student's t with $\nu=2.0$. The histogram in the bottom right corner shows distribution of the logarithmic ratio between the two SFR estimates. The offset ($\mu$), scatter ($\sigma$), and Spearman $\rho$ values are shown in the top left corner.}
\label{fig:SFRpixedfit_vs_SFRHalpha_photo_baye_uniform}
\end{figure}

The above results are obtained using the MCMC fitting method. Next, we explore the performances of fitting to the photometric SEDs using the RDSPS method with the likelihood function of Student's t with $\nu=2.0$. Figure~\ref{fig:SFRpixedfit_vs_SFRHalpha_photo_baye_uniform} shows the comparison between the SFR from the fitting and the SFR from the $H_{\alpha}$ emission. There is a good agreement between the two SFR estimates, as indicated by the small offset ($0.093$ dex), small scatter ($0.363$ dex), and high Spearman $\rho$ value ($0.710$). This result suggests that the RDSPS method with the likelihood function of Student's t with $\nu=2.0$ can give a good estimate of SFR, as good as that of the MCMC method.           

Overall, results of this analysis suggest that SED fitting to the broad-band photometry that covers FUV--NIR is capable of inferring the instantaneous SFR of a galaxy. More interestingly, it is shown in this analysis that it applies to spatially resolved ($\sim 1$ kpc) scale in the galaxy. While fitting with MCMC is computationally expensive, the RDSPS approach (which is $\sim 40$ time faster than MCMC, using the same number of cores) provides a great opportunity for an application to spatially resolved SED fitting analysis. In a future work, we will apply \verb|piXedfit| to a large sample of galaxies.
 
\section{Summary} \label{sec:summary}
In this paper, we present \verb|piXedfit|, a Python package that provides tools for analyzing the spatially resolved properties (including stellar and dust components) of galaxies from broad-band imaging data or a combination of broad-band imaging and IFS data. \verb|piXedfit| is designed to be modular, and consists of six main modules: (1) \verb|piXedfit_images| is for the image processing, (2) \verb|piXedfit_spectrophotometric| is for spatial matching between the imaging data and IFS data, (3) \verb|piXedfit_bin| is for pixel binning to maximize the $\text{S}/\text{N}$ ratio of the spatially resolved SED, (4) \verb|piXedfit_model| is for generating model SED, (5) \verb|piXedfit_fitting| is for performing SED fitting, and (6) \verb|piXedfit_analysis| is for visualization of fitting results. 

We test the capabilities of \verb|piXedfit| with two analyses in this paper: testing the SED fitting performance using mock FUV--NIR SEDs of IllustrisTNG galaxies and testing \verb|piXedfit| modules using spatially resolved spectrophotometric data of local galaxies. Overall, the testing results are summarized as follows:
\begin{enumerate}    
\item We test the performance of \verb|piXedfit_fitting| module by fitting mock FUV--NIR SEDs (photometric as well as spectrophotometric SEDs) of IllustrisTNG galaxies and then compare the inferred parameters from fitting with the true parameters. We implement various fitting approaches (the MCMC and RDSPS with likelihoods of Gaussian and Student's t with various values of $\nu$) provided within \verb|piXedfit_fitting| to compare their performances. With photometric SED that covers FUV--NIR, \verb|piXedfit_fitting| can well recover mass-weighted ages, dust optical depth, $M_{*}$, and SFR of the IllustrisTNG galaxies, for all of the fitting approaches (see Section~\ref{sec:tetsfit_illustris_TNG} and Appendix~\ref{apdx:find_dof_stdt}). The fitting to mock spectrophotometric SED improve the parameters inference, especially for the metallicity.  

\item Using the mock SEDs and SFHs of the IlustrisTNG galaxies, we test the performance of \verb|piXedfit_fitting| in inferring the SFH of a galaxy. We quantitatively assess the performance by comparing the true and inferred values of lookback times when the galaxies $M_{*}$ were only $30\%$ ($lbt_{30\% M}$), $50\%$ ($lbt_{50\% M}$), $70\%$ ($lbt_{70\% M}$), and $90\%$ ($lbt_{90\% M}$) of the current values. With FUV--NIR photometric SEDs, \verb|piXedfit_fitting| can well recover the $lbt_{30\% M}$, $lbt_{50\% M}$, $lbt_{70\% M}$, and $lbt_{90\% M}$ using all of the fitting approaches. The fitting to mock spectrophotometric SEDs improves the SFH inference.  

\item We demonstrate the performances of \verb|piXedfit| modules using spatially resolved spectrophotometric data of 20 galaxies observed by the CALIFA and MaNGA surveys. The \verb|piXedfit_images| and \verb|piXedfit_spectrophotometric| are capable of spatially matching (in resolution and sampling) of 12-bands imaging data from GALEX, SDSS, 2MASS, and WISE, and the IFS data from CALIFA and MaNGA. \verb|piXedfit_bin| is capable of binning neighboring pixels with similar SED shape and reach target $\text{S}/\text{N}$ ratios in all bands.

\item By fitting to photometric SED only, \verb|piXedfit| can predict real spectral continuum, $\text{D}_{\rm n}4000$, $H_{\alpha}$ emission, and $H_{\beta}$ emission. The residuals between the spectral continuum of the median posterior and that of the observed spectra are flat over a wide range of wavelength, in both CALIFA and MaNGA samples. The predicted $H_{\alpha}$, $H_{\beta}$, and $\text{D}_{\rm n}4000$ are consistent with the observed ones, with offsets of $0.105$, $0.131$, and $-0.024$ dex, respectively.

\item Using the $H_{\alpha}$ and $H_{\beta}$ luminosities of the observed spectra, we derive the SFR. The dust attenuation correction based on the Balmer decrement is applied. Then we compare that SFR with the SFR derived from the SED fitting. The SFR derived from the SED fitting with \verb|piXedfit_fitting| is consistent with the SFR derived from the $H_{\alpha}$ emission.               

\item While most of the fitting approaches in the \verb|piXedfit_fitting| give good inferences of stellar population properties and SFH, there are indications that the approach of RDSPS with likelihood of Student's t with $\nu \sim2$, with proper priors, can give robust (and stable) parameters inference, as good as MCMC method. With its relatively fast fitting performance ($\sim 40$ time faster than the MCMC), this fitting approach can be a good option for performing spatially resolved SED fitting for a large sample of galaxies.   
\end{enumerate}

\texttt{piXedfit} is a powerful tool for analyzing the spatially resolved properties of galaxies across wide range of resdshifts in the future era of big data in photometry from the deep and high spatial resolution multiband imaging surveys. \texttt{piXedfit} will be made publicly available on GitHub\footnote{\texttt{piXedfit} codebase: \url{https://github.com/aabdurrouf/piXedfit}}, archived in Zenodo \citep{2021Abdurrouf}, and documented at \url{https://pixedfit.readthedocs.io/en/latest/index.html}.

\acknowledgments
We thank an anonymous referee for providing useful comments that helped to improve this paper. We are grateful for supports from the Ministry of Science \& Technology of Taiwan under the grant MOST 108-2112-M-001-011, MOST 109-2112-M-001-005, and a Career Development Award from Academia Sinica (AS-CDA-106-M01). P.F.W. acknowledges the support of the fellowship from the East Asian Core Observatories Association. The computations in this research were run on the TIARA and SuMIRe clusters at ASIAA. 

This research made use of Astropy\footnote{\url{http://www.astropy.org}}, a community-developed core Python package for Astronomy \citep{2013Astropy, 2018Astropy}. This research made use of \texttt{Photutils}, an \texttt{Astropy} package for detection and photometry of astronomical sources \citep{2019bradley}. This work is based on observations made with the NASA Galaxy Evolution Explorer (GALEX), which is operated for NASA by the California Institute of Technology under NASA contract NAS5-98034. Funding for the Sloan Digital Sky Survey IV has been provided by the Alfred P. Sloan Foundation, the U.S. Department of Energy Office of Science, and the Participating Institutions. SDSS-IV acknowledges support and resources from the Center for High-Performance Computing at the University of Utah. The SDSS web site is www.sdss.org. SDSS-IV is managed by the Astrophysical Research Consortium for the Participating Institutions of the SDSS Collaboration including the Brazilian Participation Group, the Carnegie Institution for Science, Carnegie Mellon University, the Chilean Participation Group, the French Participation Group, Harvard-Smithsonian Center for Astrophysics, Instituto de Astrof\'isica de Canarias, The Johns Hopkins University, Kavli Institute for the Physics and Mathematics of the Universe (IPMU) / 
University of Tokyo, the Korean Participation Group, Lawrence Berkeley National Laboratory,  Leibniz Institut f\"ur Astrophysik Potsdam (AIP),  Max-Planck-Institut f\"ur Astronomie (MPIA Heidelberg), 
Max-Planck-Institut f\"ur Astrophysik (MPA Garching), 
Max-Planck-Institut f\"ur Extraterrestrische Physik (MPE), 
National Astronomical Observatories of China, New Mexico State University, 
New York University, University of Notre Dame, 
Observat\'ario Nacional / MCTI, The Ohio State University, 
Pennsylvania State University, Shanghai Astronomical Observatory, 
United Kingdom Participation Group,
Universidad Nacional Aut\'onoma de M\'exico, University of Arizona, 
University of Colorado Boulder, University of Oxford, University of Portsmouth, 
University of Utah, University of Virginia, University of Washington, University of Wisconsin, 
Vanderbilt University, and Yale University. This publication makes use of data products from the Two Micron All Sky Survey, which is a joint project of the University of Massachusetts and the Infrared Processing and Analysis Center/California Institute of Technology, funded by the National Aeronautics and Space Administration and the National Science Foundation. This publication makes use of data products from the Wide-field Infrared Survey Explorer, which is a joint project of the University of California, Los Angeles, and the Jet Propulsion Laboratory/California Institute of Technology, funded by the National Aeronautics and Space Administration. This study uses data provided by the Calar Alto Legacy Integral Field Area (CALIFA) survey (\url{http://califa.caha.es/}). Based on observations collected at the Centro Astronómico Hispano Alemán (CAHA) at Calar Alto, operated jointly by the Max-Planck-Institut fűr Astronomie and the Instituto de Astrofísica de Andalucía (CSIC). 


\vspace{5mm}
\facilities{GALEX, Sloan, CTIO:2MASS, FLWO:2MASS, WISE, Sloan (BOSS, MaNGA survey), CAO:3.5m (PMAS/PPAK, CALIFA survey)}


\software{\texttt{Astropy} \citep{2013Astropy},
            \texttt{Photutils} \citep{2019bradley},
            \texttt{reproject} \citep{2018robitaille},
          \texttt{SExtractor} \citep{1996bertin},
          \texttt{sewpy},
		\texttt{FSPS} \citep{2009Conroy},
		\texttt{python-FSPS} \citep{2014Foreman},
		\texttt{emcee} \citep{2013Foreman},
		\texttt{matplotlib} \citep{2007Hunter},
		\texttt{SciPy} \citep{2020Virtanen},
		\texttt{NumPy} \citep{2020Harris},
          \texttt{specutils \citep{2020nicholas}}
          }



\appendix

\section{Comparison of the Empirical PSF of the \textit{SDSS} and \textit{2MASS} with the Analytical PSF from Aniano et al. (2011)} \label{apdx:comp_empPSFs_aniano2011}
\begin{figure*}[ht]
\centering
\includegraphics[width=0.6\textwidth]{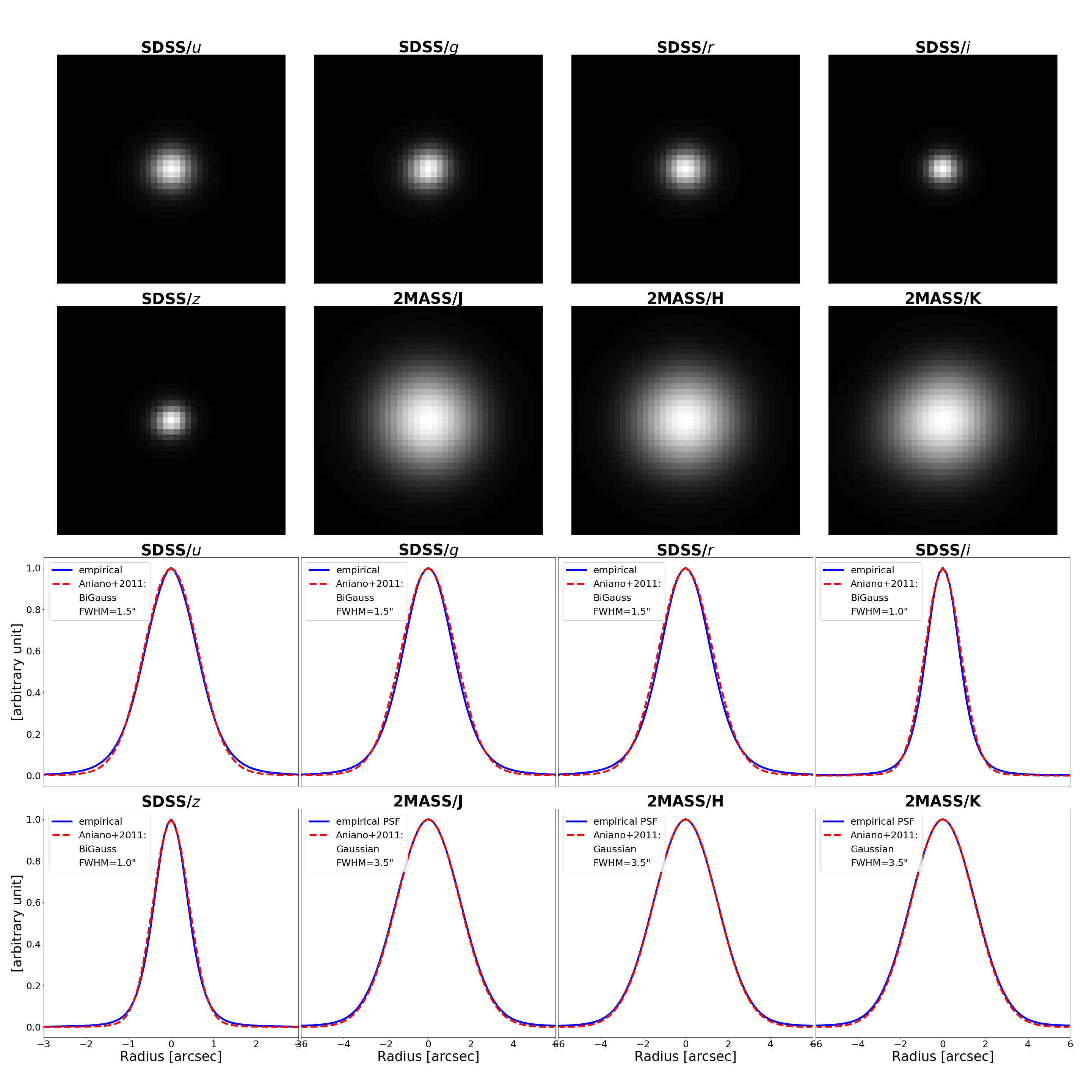}
\caption{Comparison of the empirical PSFs of the SDSS and 2MASS with the analytical PSFs from \citet{2011aniano}. In the first and second rows, empirical PSFs of the SDSS and 2MASS are shown. Comparison of those empirical PSFs with the analytical PSFs are shown in the third and fourth rows.}
\label{fig:comp_empPSF_analyticPSF}
\end{figure*}

We construct empirical PSFs of the SDSS and 2MASS using the PSF modeling functions provided by \verb|Photutils|. The \verb|Photutils| package provides tools for building an effective PSF, which can represent the net PSF of a given camera. The effective PSF is built based on the prescription in \citet{2000anderson}. First, several images of random fields are downloaded from the SDSS and 2MASS websites. Then background subtraction is done, especially for 2MASS images (the SDSS image product is background free). After that, bright stars are collected using \verb|find_peaks| function. The \verb|extract_stars| function is used to extract cutouts of the stars. Then visual inspection is done to exclude ``bad stars'', such as multiple stars in one cutout image and saturated stars. Finally, effective PSFs are constructed using \verb|EPSFBuilder| function. In building the effective PSFs, number of stars selected for $u$, $g$, $r$, $i$, $z$, $J$, $H$, and $K_{s}$ are 103, 123, 143, 170, 268, 102, 118, and 94, respectively. The constructed effective PSFs of the SDSS and 2MASS are shown in Figure~\ref{fig:comp_empPSF_analyticPSF}. We compare the empirical PSFs with analytical PSFs from \citet{2011aniano}. We found that the PSFs of $u$, $g$, and $r$ are best represented by the double Gaussian with FWHM of $1.5''$; PSFs of the $i$ and $z$ are best represented by the double Gaussian with FWHM of $1.0''$; and PSFs of 2MASS are best represented by the Gaussian with FWHM of $3.5''$. This comparison is shown in the third and fouth rows in the figure. We also construct empirical PSFs of FUV and NUV bands of GALEX using the same procedure. The constructed empirical PSFs of FUV and NUV are consistent with the PSFs from \citet{2011aniano}. The empirical PSFs from this analysis can be found at \url{https://github.com/aabdurrouf/empPSFs_GALEXSDSS2MASS}.

\section{Comparison of the performances of various fitting approaches provided in \texttt{piXedfit\_fitting} module} \label{apdx:find_dof_stdt}
In Section~\ref{sec:tetsfit_illustris_TNG}, we fit the mock FUV--NIR photometric SEDs of the TNG galaxies using \verb|piXedfit_fitting| module with 8 different fitting approaches, including the 2 posterior sampling methods (MCMC and RDSPS), the 2 likelihood functions (Gaussian and Student's t) in the RDSPS method, and 6 values of $\nu$ for the Student's t likelihood function: $0.3$, $1.0$, $2.0$, $3.0$, $5.0$, and $10.0$. The purpose of performing this fitting experiment is to compare the performances of the various fitting approaches provided within the \verb|piXedfit_fitting| module.   

In this analysis, we compare the performances of those fitting approaches in inferring 5 key parameters: $Z$, dust optical depth ($\hat{\tau}_{2}$), mass-weighted age, $M_{*}$, and SFR. Similar to what we do in Section~\ref{sec:recover_params_TNG} and~\ref{sec:recover_SFH_TNG}, for each parameter obtained with each fitting approach, we calculate the offset ($\mu$), scatter ($\sigma$), and Spearman $\rho$ coefficient of the 1D distribution of the logarithmic ratios between the inferred values from fitting and the true values. Then here, for each parameter, we compare the goodness of the recovery among the fitting approaches by directly comparing the $\mu$, $\sigma$, and $\rho$ values. 

\begin{figure*}[ht]
\centering
\includegraphics[width=0.75\textwidth]{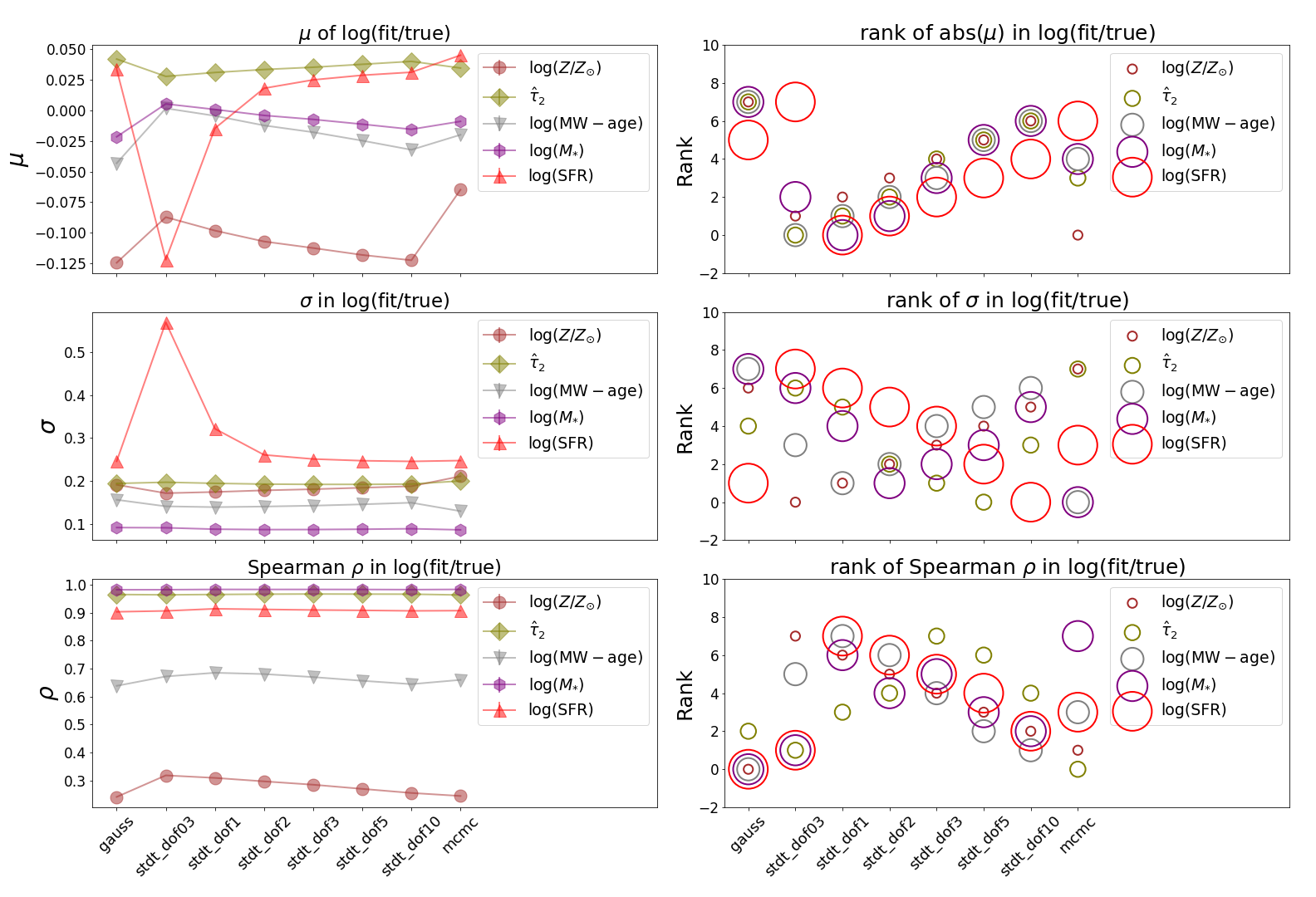}
\caption{Compilation of the $\mu$, $\sigma$, and $\rho$ for the 1D distributions of the logrithmic ratios between the inferred values from fitting to the mock photometric SEDs of TNG galaxies and the true values. In the left side, real values of $\mu$, $\sigma$, and $\rho$ associated with the parameters and the fitting approaches are shown. Different symbols represent different parameters. In the right side, for each parameter, the $\mu$, $\sigma$, and $\rho$ values associated with the various fitting approaches are sorted and ranked from smallest (ranked as 0) to highest (ranked as 7). Different parameters are shown with circles of different colors and sizes.}
 \label{fig:plot_comp_fitres_TNG}
\end{figure*}

Figure~\ref{fig:plot_comp_fitres_TNG}, left panel, shows a compilation of the values of $\mu$ (first row), $\sigma$ (second row), and Spearman $\rho$ (third row). Different fitting parameters are shown with different symbols. The horizontal axis shows the various fitting approaches. From this plots, overall, we can see that all the fitting approaches give good performances, indicated by the low values of absolute $\mu$ ($\lesssim 0.13$ dex) in all the parameters, low scatter ($\lesssim 0.3$ dex) in all the parameters, except the SFR derived with the RDSPS that uses the Student's t $\nu=0.3$ likelihood, and high $\rho$ ($\gtrsim 0.6$) in all the parameters, except $Z$. The average absolute $\mu$, $\sigma$, and $\rho$ for the [\texttt{gauss}, \texttt{stdt\_dof03}, \texttt{stdt\_dof1}, \texttt{stdt\_dof2}, \texttt{stdt\_dof3}, \texttt{stdt\_dof5}, \texttt{stdt\_dof10}, \texttt{mcmc}] are [$0.0530$, $0.0489$, $0.0298$, $0.0350$, $0.0396$, $0.0441$, $0.0483$, $0.0347$], [$0.1765$, $0.2344$, $0.1841$, $0.1725$, $0.1714$, $0.1720$, $0.1735$, $0.1755$], and [$0.7460$, $0.7685$, $0.7714$, $0.7677$, $0.7629$, $0.7569$, $0.7510$, $0.7516$], respectively.

For the ease of comparison among the fitting approaches, in the right panel of Figure~\ref{fig:plot_comp_fitres_TNG}, the $\mu$, $\sigma$, and $\rho$ values associated with each parameter are sorted and ranked from smallest (ranked as $0$) to highest (ranked as 7). For the $\mu$, the absolute value is considered. Different parameters are shown with circles of different colors and sizes. This plot indicates that the RDSPS method that uses Student's t likelihood function with $\nu\sim1.0-3.0$ can possibly outperform the other fitting approaches. However, more fitting experiments, such as those using more realistic panchromatic mock SEDs or comparisons between the inferred parameters from fitting with other empirical independent indicators are needed to verify this finding. A better inference given by the Student's t likelihood over the Gaussian one is possibly caused by the heavier tails of the Student's t function which can better accomodate all the model SEDs (despite having large $\chi^{2}$) in the Bayesian inference process compared to the Gaussian function. 

\bibliography{pixedfit}{}
\bibliographystyle{aasjournal}



\end{document}